\documentclass[journal]{IEEEtran}
\usepackage{citesort,amssymb,amsthm,amsmath,graphicx}
\graphicspath{{Figure//}}

\theoremstyle{definition}
\newtheorem{MAX-PTX-NAE}{Proposition}
\newtheorem{MAX-RS-AE}[MAX-PTX-NAE]{Proposition}

\begin{document}
\title{On the Design of Artificial-Noise-Aided\\Secure Multi-Antenna Transmission\\in Slow Fading Channels}
\author{Xi Zhang, \IEEEmembership{Student Member, IEEE,} Xiangyun Zhou, \IEEEmembership{Member, IEEE,}\\ and Matthew R. McKay, \IEEEmembership{Member, IEEE}
\thanks{Copyright (c) 2012 IEEE. Personal use of this material is permitted. However, permission to use this material for any other purposes must be obtained from the IEEE by sending a request to pubs-permissions@ieee.org.}
\thanks{X.~Zhang and M.~R.~McKay are with the Department of Electronic and Computer Engineering, Hong Kong University of Science and Technology, Hong Kong (e-mails: xizhangx@ust.hk, eemckay@ust.hk).}
\thanks{X. Zhou is with the Research School of Engineering, the Australian National University, Australia (e-mail: xiangyun.zhou@anu.edu.au).}
\thanks{The work of X.~Zhang and M.~R.~McKay was supported by the Hong Kong Research Grants Council (Grant No. 616312).}
\thanks{The work of X.~Zhou was supported by the Australian Research Council's Discovery Projects funding scheme (Project No. DP110102548).}
}
\maketitle

\begin{abstract}
In this paper, we investigate the design of artificial-noise-aided secure multi-antenna transmission in slow fading channels. The primary design concerns include the transmit power allocation and the rate parameters of the wiretap code. We consider two scenarios with different complexity levels: i) the design parameters are chosen to be fixed for all transmissions, ii) they are adaptively adjusted based on the instantaneous channel feedback from the intended receiver. In both scenarios, we provide explicit design solutions for achieving the maximal throughput subject to a secrecy constraint, given by a maximum allowable secrecy outage probability. We then derive accurate approximations for the maximal throughput in both scenarios in the high signal-to-noise ratio region, and give new insights into the additional power cost for achieving a higher security level, whilst maintaining a specified target throughput. In the end, the throughput gain of adaptive transmission over non-adaptive transmission is also quantified and analyzed.
\end{abstract}


\section{Introduction}\label{Introduction}
\IEEEPARstart{W}{ith} the rapid development of wireless technology, an ever-increasing amount of sensitive data (e.g., private conversation, credit card information) is transmitted over wireless networks. However, the broadcasting nature of the wireless medium makes it especially vulnerable to malicious interception. Currently, cryptographic algorithms, typically designed without exploiting the physical properties of the wireless medium, are used to keep broadcasted messages confidential. For such techniques, although the expenditure of interception may be very high, providing robust encryption algorithms is becoming ever more challenging, due to the continuing development of computing devices. By taking the physical properties of the wireless channels into consideration, the recently developed physical-layer security techniques can guarantee secure transmission regardless of the eavesdropper's computational capability. As such, these techniques have drawn a lot of recent attention from the research community.

\subsection{Background and Previous Work}
The notion of perfect secrecy was first introduced by Shannon~\cite{Shann1949}. Subsequently, pioneering works on physical-layer security~\cite{Wyner1975,Csiszar1978} proved that there exist coding schemes which can ensure transmission reliability and perfect secrecy simultaneously. Many recent papers have expanded upon these initial contributions, considering different system configurations and assumptions. Particularly, multi-antenna techniques have been extensively studied as a means for achieving security enhancements~\cite{Li2007,Shafiee2007,Oggier2008,Liu2010,Khisti20102}. However, in much of the literature on physical-layer security, the eavesdropper's channel state information (CSI) was assumed to be available at the transmitter, which is usually impractical. To relax this strong assumption, the authors in~\cite{Goel2008} proposed a multi-antenna transmission scheme that inserts artificial noise into the transmitted signal in a controlled manner, thus to confuse the malicious eavesdropper. This transmission scheme requires the instantaneous CSI feedback from the intended receiver, but not the eavesdropper, which is a major advancement toward practical secure communications.

Building on the ideas from~\cite{Goel2008}, the design and analysis of artificial-noise-aided transmission has been further studied for both fast and slow fading channels~\cite{Zhou2010,Gerbracht2010,Lin2011,Ng2011,GhoGho2011,Li2011,Gerbracht2012}. For fast fading channels, the channel coherence time is much shorter than the codeword length and the ergodic secrecy rate is often used as the performance metric for designing beamforming and power allocation strategies~\cite{Zhou2010,Gerbracht2010,Lin2011}. For slow fading channels, the channel coherence time is usually longer than the codeword length, and in such scenarios outage-based formulations become more appropriate. To this end, various secrecy outage formulations were proposed first in~\cite{Parada2005,Bloch2008} and recently in~\cite{Yuksel2011,Zhou2011}. In particular, the first formulation developed in~\cite{Parada2005,Bloch2008} has been used for studying artificial-noise-aided multi-antenna transmission schemes in~\cite{Ng2011,GhoGho2011,Li2011,Gerbracht2012}. This secrecy outage formulation characterizes the possibility of having a secure and reliable transmission, without distinguishing secrecy from reliability. In other words, a secrecy outage event defined therein may occur due to either an insecure link to the eavesdropper or an unreliable link to the intended receiver. To better assist the secure transmission design, revised secrecy outage formulations were independently developed in~\cite{Yuksel2011} and~\cite{Zhou2011}. In these revised outage formulations, a secrecy outage event arises solely due to an insecure link to the eavesdropper; thus, the secrecy and reliability performance can be measured separately. These revised secrecy outage formulations can be utilized to obtain a better understanding and more practically-oriented designs for the artificial-noise-aided secure multi-antenna transmission.

\subsection{Our Approach and Contributions}
In this paper, we provide new design guidelines for artificial-noise-aided secure multi-antenna transmission in slow fading channels, based on the recently developed secrecy outage formulation in~\cite{Zhou2011}. This formulation allows us to measure the secrecy and reliability performance for any given rate parameters of the wiretap code. In turn, we are able to set the rate parameters to achieve a target secrecy level, given by a maximum allowable secrecy outage probability. To the best of the authors' knowledge, no prior work on artificial-noise-aided multi-antenna transmission has considered the rate parameters of the wiretap code as design parameters.

Our main contributions include explicit design solutions and new performance analysis results for the throughput-maximizing transmission schemes with either fixed-rate or adaptive-rate encoder, under a constraint on the level of secrecy. The design concerns include the rate parameters of the wiretap code, as well as the transmit power allocation between the information-bearing signal and the artificial noise. We consider two scenarios with different system complexities:

\begin{itemize}
\item In the first scenario, the design parameters are optimized off-line and remain fixed for all transmissions. We divide our design into two steps: The first step minimizes the transmission delay for a given data rate, while the second step maximizes the average throughput. In the first step, closed-form solutions are derived for the optimal system parameters and the secrecy-delay tradeoff for fixed-rate transmission is captured. In the second step, the average throughput is maximized by numerically optimizing the data rate with the optimal designs of all other system parameters already given in closed form. To obtain further analytical insights, we derive a high signal-to-noise ratio (SNR) approximation of the optimal data rate to maximize the average throughput. Focusing on the high SNR region, we further investigate the additional power cost incurred by imposing or strengthening the secrecy constraint, while guaranteeing a specified target throughput.
\item In the second scenario, the design parameters are dynamically adjusted based on the instantaneous channel feedback from the intended receiver. We provide an analytical solution to the optimal system parameters that maximize the achievable data rate for each realization of the intended channel, under the secrecy constraint, such that the average throughput is also maximized. In the high SNR region, we present accurate approximations for the maximal throughput, which enable us to study the additional power cost incurred by imposing or strengthening the secrecy constraint, while achieving a specified target throughput. Finally, we analyze the throughput gain of adaptive-rate transmission over fixed-rate transmission. Whilst doing adaptation is always beneficial in terms of increasing the throughput, our analysis shows that in the high SNR region, the throughput gain is most significant when the number of transmit antennas is small. Moreover, we show that the throughput gain brought by doing adaption is not very sensitive to the required secrecy level.
\end{itemize}

We point out that, similar to our work, a very recent contribution~\cite{Romero-Zurita2012} also used a secrecy outage formulation along the lines of~\cite{Yuksel2011,Zhou2011} to study artificial-noise-aided multi-antenna transmission. The main focus therein was to minimize the power consumption for given levels of secrecy and quality-of-service performance, while the transmission rates were not part of the design consideration. In contrast, we consider a transmission system with a fixed transmit power, and optimize the rate parameters of the wiretap code as well as the transmit power allocation to maximize the average throughput, subject to (s.t.) a secrecy outage constraint.

\section{System Model and Performance Metric}
We consider the transmission from Alice to Bob in the presence of an eavesdropper Eve. Alice is equipped with multiple transmit antennas ($N\geq2$) while Bob and Eve each has one receive antenna. Thus, the channel from Alice to Bob is multiple-input and single-output (MISO). We assume a non-line-of-sight rich scattering environment, and as such, model all channels as uncorrelated Rayleigh fading. It is also assumed that Bob can estimate his channel accurately and use a perfect feedback link to inform Alice about his instantaneous CSI. This feedback link is not secure and can be intercepted by Eve. We further assume that the coherence time is long enough to support the wiretap code~\cite{Wyner1975} and the time used for learning the channel and feeding back the CSI is negligible. Assuming Eve is a passive eavesdropper, the instantaneous CSI of Eve is thereby unavailable to Alice.

The $N$ dimensional symbol vector to be transmitted is defined as ${\bf x}$ and the received signal at Bob is given by
\begin{align}
y_b={\bf h}^{T}{\bf x}+n_b\;,
\label{eq:Yb-ORI-NAE}
\end{align}
where the $N\times1$ vector ${\bf h}$ is the channel fading gain from Alice to Bob and $n_b$ is the receiver noise at Bob. The entries of ${\bf h}$ and $n_b$ are assumed to be independent and identically distributed~(i.i.d.) zero-mean complex Gaussian variables with unit variance.

Similarly, the received signal at Eve is given by
\begin{align}
y_e={\bf g}^{T}{\bf x}+n_e\;,
\label{eq:Ye-ORI-NAE}
\end{align}
where the $N\times1$ vector ${\bf g}$ captures the channel fading gain from Alice to Eve and $n_e$ is the receiver noise at Eve. The entries of ${\bf g}$ are assumed to be i.i.d. zero-mean complex Gaussian variables each with variance $\sigma_g^2$.

\subsection{Transmit Beamforming with Artificial Noise Generation}
The authors in~\cite{Goel2008} introduced the concept of generating artificial noise to guarantee secure transmission. The key idea is outlined as follows. Alice generates an orthonormal basis of $\mathbb{C}^{N}$ as ${\bf W}=\left[{\bf w}_1\;\;{\bf W}_2\right]$, where ${\bf w}_1={\bf h}^*/{||{\bf h}||}$. Then she mixes some artificial noise with the message symbol $u$ as
\begin{align}
{\bf x}={\bf w}_1u+{\bf W}_2{\bf v}\;,\nonumber
\end{align}
where $u$ is complex Gaussian distributed and ${\bf v}$ is the artificial noise vector.

With this beamforming strategy, by~(\ref{eq:Yb-ORI-NAE}) and~(\ref{eq:Ye-ORI-NAE}), the received signal at Bob becomes
\begin{align}
y_b={\bf h}^{T}{\bf w}_1u+{\bf h}^{T}{\bf W}_2{\bf v}+n_b=||{\bf h}||u+n_b\;,
\label{eq:Yb-BEA-NAE}
\end{align}
while the received signal at Eve becomes
\begin{align}
y_e={\bf g}^{T}{\bf w}_1u+{\bf g}^{T}{\bf W}_2{\bf v}+n_e=g_1u+{\bf g}_2^{T}{\bf v}+n_e\;,
\label{eq:Ye-BEA-NAE}
\end{align}
where $g_1={\bf g}^{T}{\bf w}_1$ and ${\bf g}_2^T={\bf g}^{T}{\bf W}_2$.

The total transmit power available at Alice is denoted as~$P$. The power allocation ratio $\Phi$ is defined as the fraction of the information-bearing signal power to the total transmit power. Thus, the variance of $u$ is set to $P\Phi$. By~(\ref{eq:Yb-BEA-NAE}), it is clear that Alice is performing maximal ratio transmission on the channel to Bob and the instantaneous effective channel gain is $||{\bf h}||^2$. Therefore, the instantaneous channel capacity to Bob is~$C_b=\log_2\left(1+P\Phi{||{\bf h}||}^2\right)$, where \mbox{${||{\bf h}||}^2\sim{\rm Gamma}\left(N,1\right)$}. The complementary cumulative distribution function (c.c.d.f.) of~$||{\bf h}||^2$ is $\bar{F}_{||{\bf h}||^2}(r)={\Tilde{\Gamma}\left(N,r\right)}$, where $\Tilde{\Gamma}\left(\cdot,\cdot\right)$ is the regularized upper incomplete gamma function. All transmit power not allocated to $u$ is equally assigned to the $N-1$ entries of ${\bf v}$, due to the absence of Eve's CSI. Therefore, the entries of ${\bf v}$ are set to be i.i.d. zero-mean complex Gaussian variables each with variance $\sigma_v^2=\frac{P\left(1-\Phi\right)}{N-1}$. The power allocation ratio $\Phi$ may be dynamically adjusted based on ${\bf h}$.

Since the noise power at Eve is typically unknown to Alice, a robust approach, as done in~\cite{Zhou2010}, is to design for the worst-case scenario and assume that there is no receiver noise at Eve (i.e.,~$n_e=0$). Therefore, from~(\ref{eq:Ye-BEA-NAE}), for a given realization of the channel fading gains (i.e.,~${\bf h}$ and~${\bf g}$), the instantaneous SNR at Eve is given by
\begin{align}
\gamma_e=\frac{|g_1|^2\sigma_u^2}{||{\bf g}_2||^2\sigma_v^2}=\frac{N-1}{\Phi^{-1}-1}\frac{|g_1|^2}{||{\bf g}_2||^2}\;.\nonumber
\end{align}
Since ${\bf g}$ has i.i.d. complex Gaussian entries each with variance $\sigma_g^2$ and ${\bf W}$ is unitary, \mbox{${\bf g}^T{\bf W}=[g_1\;{\bf g}_2]$} also has i.i.d. complex Gaussian entries each with variance $\sigma_g^2$. Using~\mbox{\cite[Eq. 19]{Gao1998}}, the c.c.d.f. of $\gamma_e$ can be characterized as
\begin{align}
\bar{F}_{\gamma_e}\left(\gamma_e\right)=\left(1+\gamma_e{\left({\frac{\Phi^{-1}-1}{N-1}}\right)}\right)^{1-N}\;.
\label{eq:CCDF-GE-NAE}
\end{align}
As $N\rightarrow\infty$, this converges to the c.c.d.f. of an exponential distribution with mean $\frac{\Phi}{1-\Phi}$. Note that if $\Phi$ is adaptively adjusted based on ${\bf h}$, the distribution of the received SNR at Eve would be changed dynamically.

\subsection{Secure Transmission and Throughput}
Using the well-known wiretap code~\cite{Wyner1975}, data is encoded before transmission. The rates of the transmitted codeword and the secret message are denoted as $R_b$ and $R_s$, respectively. In this paper, we refer to the rates $R_b$ and $R_s$ as the rate parameters of the wiretap code. The rate redundancy \mbox{$R_e\triangleq{R_b-R_s}$} is intentionally added in order to provide secrecy against eavesdropping. If the instantaneous channel capacity to Eve $C_e=\log_2\left(1+\gamma_e\right)$ is larger than $R_e$, perfect secrecy cannot be achieved and a secrecy outage is deemed to occur.

We consider an on-off transmission scheme~\cite{Zhou2011} where Alice decides to transmit or not, based on her knowledge of Bob's channel ${\bf h}$. To be specific, Alice transmits only if the effective channel gain $||{\bf h}||^2$ exceeds a threshold $\mu$. As will be seen, this on-off scheme is adopted to prevent undesirable transmissions which incur capacity outages (i.e., $R_b>C_b$) or unacceptably high risk of secrecy outage (which occurs when $C_e>R_e$). With the on-off transmission strategy, the transmit probability is
\begin{align}
p_{\rm tx}\left(\mu\right)=\bar{F}_{||{\bf h}||^2}\left(\mu\right)=\Tilde{\Gamma}\left(N,\mu\right)\;.\label{eq:PTX-NAE}
\end{align}
Since the channel fading gain changes from time to time, Alice would transmit once the effective channel gain exceeds the preselected threshold (i.e., $||{\bf h}||^2>\mu$). In this way, the value of the transmit probability $p_{\rm tx}$ is directly related to the average delay, i.e., the larger $p_{\rm tx}$, the shorter the expected delay.

Since the rates of the wiretap code may be adaptively adjusted based on Bob's channel feedback, the rates $R_b$ and $R_s$ are potentially functions of ${\bf h}$. The average throughput is then defined as
\begin{align}
\eta&=\mathbb{E}_{\bf h}\left[R_s\left({\bf h}\right)\right]\;\;{\rm (bits/channel\; use)}\;,\label{eq:THR-DEF}
\end{align}
where $R_s\left({\bf h}\right)=0$ for $||{\bf h}||^2\leq\mu$, as a consequence of the on-off transmission protocol. Note that this throughput definition is meaningful only when: i) the risk of secrecy outage is under control, ii) the intended receiver can decode the messages correctly. As will be seen later, our designs can satisfy these two conditions simultaneously; thus, it is meaningful to apply this throughput definition.

\subsection{Secrecy Performance Characterization}\label{Secrecy Performance Characterization}
From~(\ref{eq:CCDF-GE-NAE}), it is clear that changing the power allocation would effectively alter the distribution of the received SNR at Eve. With on-off transmission, for a given ${\bf h}$, if ${||{\bf h}||^2}>\mu$, Alice would specify three parameters: $R_b\left({\bf h}\right)$, $R_s\left({\bf h}\right)$ and $\Phi\left({\bf h}\right)$ for transmission; otherwise, she stops transmission. Therefore, the secrecy outage probability is given by
\begin{align}
&{p_{\rm so}}\left(R_b\left({\bf h}\right),R_s\left({\bf h}\right),\Phi\left({\bf h}\right)\right)\nonumber\\
=&\begin{cases}
\Pr\left(C_e\left(\Phi\left({\bf h}\right)\right)>R_b\left({\bf h}\right)-R_s\left({\bf h}\right)\right)\;&{\rm if}\;||{\bf h}||^2>\mu\;,\\
0\;&{\rm other}\;,
\end{cases}\nonumber
\end{align}
where
\begin{align}
&\Pr\left(C_e\left(\Phi\left({\bf h}\right)\right)>R_b\left({\bf h}\right)-R_s\left({\bf h}\right)\right)\nonumber\\
=&\left(1+\left(2^{R_b\left({\bf h}\right)-R_s\left({\bf h}\right)}-1\right){\left({\frac{\Phi^{-1}\left({\bf h}\right)-1}{N-1}}\right)}\right)^{1-N}\;,\label{eq:INS-SOPH}
\end{align}
evaluated from~(\ref{eq:CCDF-GE-NAE}).

Since we assumed that $n_e=0$, the above-mentioned $p_{\rm so}$ is an upper bound on the actual secrecy outage probability. It depends on the power allocation ratio $\Phi$, but not the transmit power $P$.

The overall secrecy outage probability of this transmission system is given by
\begin{align}
\bar{p}_{\rm so}=\mathbb{E}\left[p_{\rm so}\left(R_b\left({\bf h}\right),R_s\left({\bf h}\right),\Phi\left({\bf h}\right)\right)|\,||{\bf h}||^2>\mu\right]\;.\label{eq:AVG-SOP}
\end{align}

In sequel, we consider the optimization of the throughput performance, given a maximum allowable secrecy outage probability $\epsilon$.

\section{Secure Transmission Design\\with Non-Adaptive Encoder}\label{Secure Transmission Design with Non-Adaptive Encoder}
In this section, we consider the scenario where a single codebook is used by Alice and Bob, thus $R_b$ and $R_s$ are carefully chosen to have fixed values which do not change with ${\bf h}$. In addition, the value of $\Phi$ is also optimized off-line and remains fixed. We refer to this as the non-adaptive encoding (NAE) scheme.

The primary design objective is to meet the secrecy requirement given on the secrecy outage probability. For the NAE scheme, the design parameters $R_b$, $R_s$ and $\Phi$ have fixed values for all transmissions (i.e., they do not change with ${\bf h}$); thus, (\ref{eq:INS-SOPH}) becomes independent of ${\bf h}$. Consequently, the conditional expectation in~(\ref{eq:AVG-SOP}) can be ignored, and the overall secrecy outage probability $\bar{p}_{\rm so}$ in this case is described by~(\ref{eq:INS-SOPH}). In other words, for the NAE scheme, the overall secrecy outage probability is a function of $R_b$, $R_s$ and $\Phi$, and the secrecy requirement can be written as $\bar{p}_{\rm so}\left(R_b,R_s,\Phi\right)\leq\epsilon$.

Meanwhile, as already mentioned, on-off transmission with threshold $\mu$ is adopted to prevent undesirable transmissions which incur capacity outages at Bob. When Alice decides to transmit (i.e., $||{\bf h}||^2>\mu$), the channel capacity to Bob is
\begin{align}
{C_b}=\log_2\left(1+P\Phi||{\bf h}||^2\right)>\log_2\left(1+P\Phi\mu\right)\;.\nonumber
\end{align}
Thus, with $R_b$ satisfying \mbox{$R_b\leq\log_2\left(1+P\Phi\mu\right)$}, the transmitted messages can be decoded at Bob with an arbitrarily low error probability.

From~(\ref{eq:PTX-NAE}) and~(\ref{eq:THR-DEF}), the throughput of the NAE scheme is given by
\begin{align}
\eta_{\rm NAE}={p_{\rm tx}\left(\mu\right)}\times{R_s}\;.\label{eq:Thr_Def_NAE}
\end{align}
Now we consider throughput optimization of this NAE scheme under the secrecy constraint. Combining the above discussions, the throughput optimization problem is posed as follows:
\begin{align}
\mathop{\max}_{\mu,R_b,R_s,\Phi}\;\;&{p_{\rm tx}}\left(\mu\right)\times{R_s}\nonumber\\
{\rm s.t.}\;\;&\bar{p}_{\rm so}\left(R_b,R_s,\Phi\right)\leq\epsilon\;,\nonumber\\
&{R_b}\leq\log_2\left(1+P\Phi\mu\right)\;,\nonumber\\
&0\leq{R_s}\leq{R_b}\;,\nonumber\\
&0\leq\Phi\leq1\;,\nonumber\\
&0\leq\mu\;.\label{eq:Thr-Max-NAE}
\end{align}
We solve the problem in two steps. In the first step, we fix $R_s$ and maximize $p_{\rm tx}\left(\mu\right)$ with respect to (w.r.t.) $\mu$, $R_b$ and $\Phi$. Since $R_s$ is fixed, by maximizing $p_{\rm tx}$, this step can be viewed as a delay-minimizing design, while satisfying the secrecy constraint. With a minor abuse of notation, the maximal $p_{\rm tx}$ for a given $R_s$ is denoted as $p_{\rm tx}^{\max}\left(R_s\right)$, where $\mu$ is optimized already. Then, in the second step, we consider the throughput maximization, i.e., maximizing $p_{\rm tx}^{\max}\left(R_s\right)\times{R_s}$ w.r.t. $R_s$.

\subsection{Delay Minimization}\label{Delay Minimization}
From~(\ref{eq:Thr-Max-NAE}), the delay minimization problem is posed as
\begin{align}
\mathop{\max}_{\mu,R_b,\Phi}\;\;&{p_{\rm tx}}\left(\mu\right)\nonumber\\
{\rm s.t.}\;\;&\bar{p}_{\rm so}\left(R_b,\Phi\right)\leq\epsilon\;,\nonumber\\
&{R_b}\leq\log_2\left(1+P\Phi\mu\right)\;,\nonumber\\
&{R_s}\leq{R_b}\;,\nonumber\\
&0\leq\Phi\leq1\;,\nonumber\\
&0\leq\mu\;.\label{eq:PTX-OPT-SEC-NAE}
\end{align}
To simplify the expressions throughout the paper, we define the quantity:
\begin{align}
\lambda\left(\epsilon,N\right):=\left(N-1\right)\left(\epsilon^{\frac{1}{1-N}}-1\right)\;.\label{eq:lambda}
\end{align}
The solution to~(\ref{eq:PTX-OPT-SEC-NAE}) is presented as follows, while the derivation is relegated to Appendix~\ref{Proof for LBL-MAX-PTX-NAE}.

\begin{MAX-PTX-NAE}\label{LBL-MAX-PTX-NAE}
With a non-adaptive encoder, for a prescribed message rate $R_s$ and under the secrecy constraint $\bar{p}_{\rm so}\leq\epsilon$, the maximal transmit probability in~(\ref{eq:PTX-OPT-SEC-NAE}) leading to the minimal average delay is given by:
\begin{align}
p_{\mathrm{tx}}^{\max}&=\Tilde{\Gamma}\left(N,\frac{1}{P}\left(\sqrt{2^{R_s}\lambda\left(\epsilon,N\right)}+\sqrt{2^{R_s}-1}\right)^2\right)\;.
\label{eq:PTX-MAX-NAE}
\end{align}
This is achieved with the following choice of system parameters:
\begin{align}
\Phi^{*}&=\frac{\sqrt{2^{R_s}-1}}{\sqrt{2^{R_s}\lambda\left(\epsilon,N\right)}+\sqrt{2^{R_s}-1}}\;,\label{eq:OPT-PHI-MAX-PTX-NAE}\\
R_b^{*}&=R_s+\log_2\left(\sqrt{\left(1-2^{-R_s}\right)\lambda\left(\epsilon,N\right)}+1\right)\;,\label{eq:OPT-RB-MAX-PTX-NAE}\\
\mu^{*}&=\frac{2^{R_b^{*}}-1}{P\Phi^*}\;.\nonumber
\end{align}
\end{MAX-PTX-NAE}

From~(\ref{eq:OPT-PHI-MAX-PTX-NAE}) and~(\ref{eq:OPT-RB-MAX-PTX-NAE}), several observations can be made:
\begin{itemize}
\item $R_b^*$ and $\Phi^*$ are independent of $P$. This is explained by noting that with the zero-noise assumption of Eve, the secrecy outage probability depends only on $R_b$ and $\Phi$. Therefore, for a given $\Phi$, whether the secrecy constraint can be satisfied depends only on $R_b$. Since the latter optimization w.r.t. $\Phi$ also does not involve $P$, the independence is then observed.
\item $\Phi^{*}$ increases with $N$ (since $\lambda\left(\epsilon,N\right)$ decreases with $N$, as shown in Appendix~\ref{Proof for Monotonicity of Quantity lambda}), approaching the constant:
    \begin{align}
    \lim_{N\rightarrow\infty}{\Phi^{*}}=\frac{1}{1+\sqrt{{\ln\left(\frac{1}{\epsilon}\right)}/\left({1-2^{-R_s}}\right)}}\;.\nonumber
    \end{align}
    This result follows the intuition that when more transmit antennas are installed, a smaller fraction of power can be used for generating artificial noise while still ensuring the required security level.
\end{itemize}

\begin{figure}[htbp]
\centering
\includegraphics[width=0.9\linewidth]{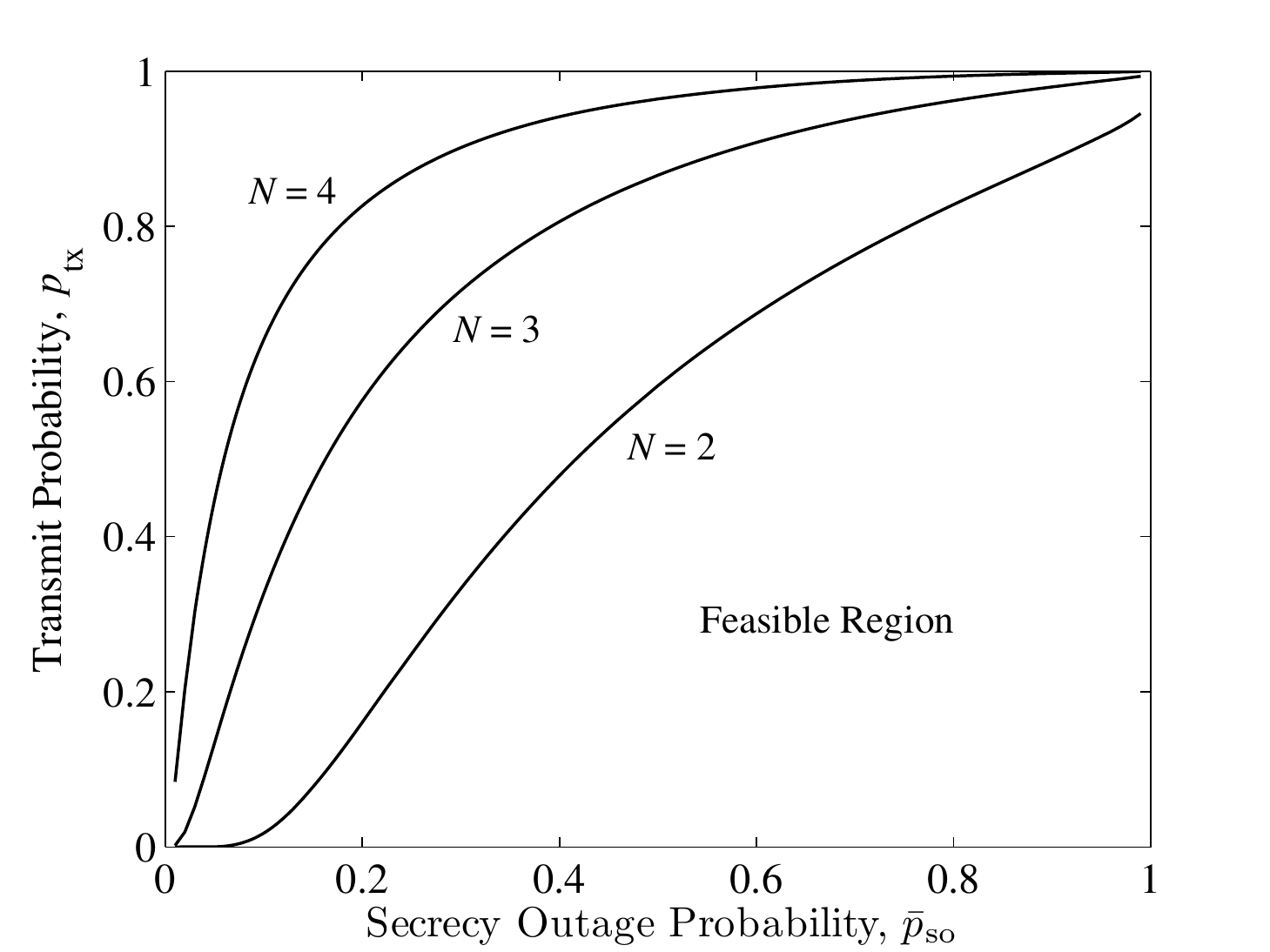}
\caption{Optimal tradeoff between the secrecy and delay performance of the NAE scheme for different numbers of transmit antennas, with $P=10$ dB and $R_s=2$ bits/channel use.}
\label{fig:Opt_Tdf_N_NAE}
\end{figure}

\begin{figure}[htbp]
\centering
\includegraphics[width=0.9\linewidth]{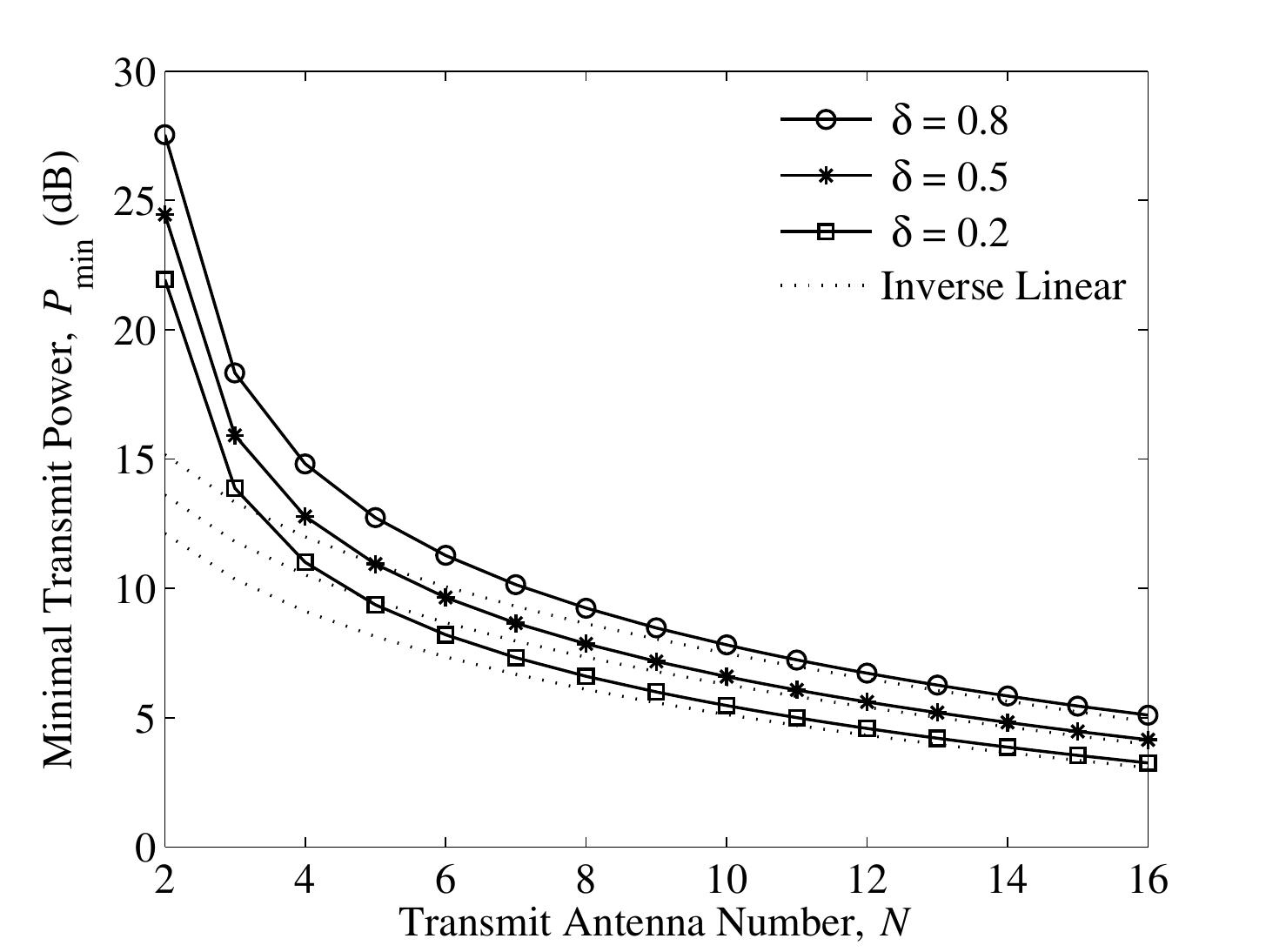}
\caption{Minimal power consumption of the NAE scheme under joint secrecy and delay constraint versus the number of transmit antennas for different delay constraints, with $\epsilon=0.01$ and $R_s=2$ bits/channel use.}
\label{fig:Pmin_N_Del_NAE}
\end{figure}

The points above focus on the optimal system parameters. Some useful insights regarding the system performance and design implications may also be concluded:
\begin{itemize}
\item For a given transmission system (i.e., given $N$ and $P$) and a prescribed message rate $R_s$, the optimal tradeoff curve between the secrecy and delay performance is completely specified by~(\ref{eq:PTX-MAX-NAE}). This is observed by noting that $p_{\rm tx}^{\max}$ is achieved when $\bar{p}_{\rm so}=\epsilon$, while varying $\epsilon$ effectively traces the optimal tradeoff. In~\cite{Zhang2011A}, we considered the minimization of the overall secrecy outage probability under an average delay constraint (i.e., \mbox{$\min\;\bar{p}_{  \rm so}\;{\rm s.t.}\;p_{\rm tx}\geq\delta$}), which is the symmetric problem to~(\ref{eq:PTX-OPT-SEC-NAE}). Since the tradeoff curve obtained in~\cite{Zhang2011A} coincides with~(\ref{eq:PTX-MAX-NAE}), we conclude that each point on the optimal tradeoff curve is Pareto optimal~\cite{Boyd2004}. As shown in Fig.~\ref{fig:Opt_Tdf_N_NAE}, the area under each curve represents the feasible region of $\left(\bar{p}_{\rm so},p_{\rm tx}\right)$, points of which are achievable.
\item If the system is designed to operate under a joint secrecy and delay constraint (i.e., $\bar{p}_{\rm so}\leq\epsilon$ and ${p_{\rm tx}}\geq\delta$), the minimum required transmit power $P_{\min}$ can be computed by replacing~$p_{\rm tx}^{\max}$ in~(\ref{eq:PTX-MAX-NAE}) with $\delta$ and solving the obtained equality w.r.t. $P$. This comes from the Pareto optimality indicated above and the fact that increasing $P$ would effectively expand the feasible region. If extra antennas may be added, the decreasing rate of $P_{\min}$ is most significant for small $N$. For the special case $\delta=0.5$, it can be proved that $P_{\min}$ decreases inverse linearly as $N$ goes large, using~\cite[Eq. 6.2]{Temme1992}. As shown in Fig.~\ref{fig:Pmin_N_Del_NAE}, this inverse linear behavior holds for a wide range of $\delta$ (here, the dotted lines are properly scaled inverse linear curves, shown for reference).
\end{itemize}

\subsection{Throughput Optimization and Analysis}\label{Throughput Optimization and Analysis}
Having maximized $p_{\mathrm{tx}}$ for a given $R_s$, from~(\ref{eq:Thr-Max-NAE}) and~(\ref{eq:PTX-MAX-NAE}), the optimal message rate that maximizes the throughput is given by
\begin{align}
R_s^*=\mathop{\arg\max}_{R_s>0}\;{\eta}_{\rm NAE}=\mathop{\arg\max}_{R_s>0}\;p_{\mathrm{tx}}^{\max}\left(R_s\right)\times{R_s}\;,\label{eq:THR-OPT-RS-NAE}
\end{align}
and the corresponding maximal throughput is given by
\begin{align}
\eta_{\rm NAE}^*={p_{\rm tx}^{\max}\left(R_s^*\right)}\times{R_s^*}\;.\label{eq:Thr_Max_NAE}
\end{align}

\begin{figure}[htbp]
\centering
\includegraphics[width=0.9\linewidth]{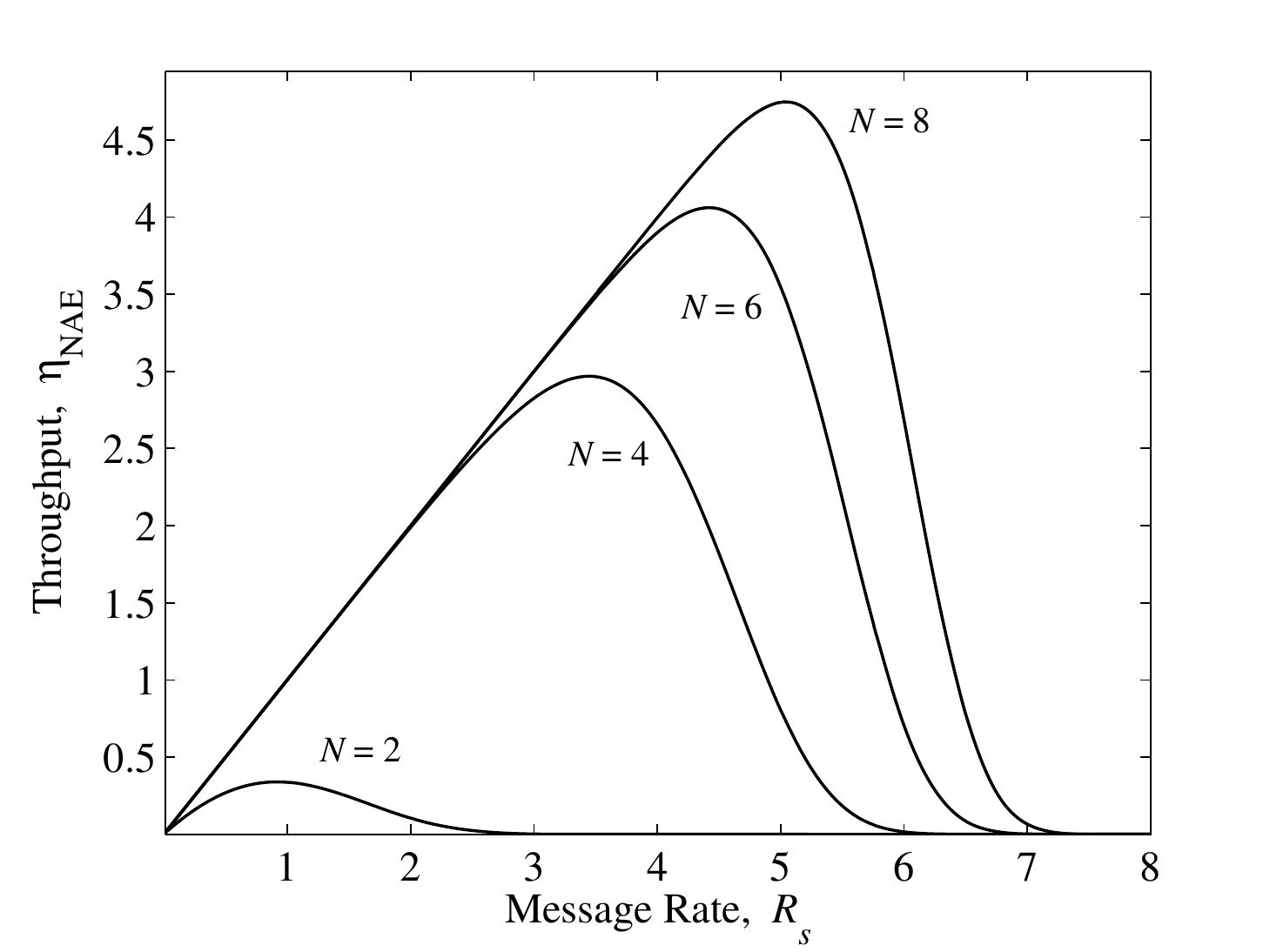}
\caption{Throughput of the NAE scheme versus the message rate for different numbers of transmit antennas, with $P=20$ dB and $\epsilon=0.01$.} \label{fig:Thr_Rs_NAE}
\end{figure}

From~(\ref{eq:PTX-MAX-NAE}) and~(\ref{eq:THR-OPT-RS-NAE}), it can be shown that if $R_s$ is too small, even though $p_{\rm tx}$ may be high (i.e., close to one), the value of $\eta_{\rm NAE}$ is still very small; whereas, if $R_s$ is too large, the value of $p_{\rm tx}$ and therefore $\eta_{\rm NAE}$ will also become very small. Moreover, by differentiating ${\eta}_{\rm NAE}=p_{\mathrm{tx}}^{\max}\left(R_s\right)\times{R_s}$ w.r.t. $R_s$, one can verify that the derivative is first positive and then negative with increasing $R_s$; thus $R_s^*$ is unique. As such, even if the objective function is non-concave, as shown in Fig.~\ref{fig:Thr_Rs_NAE}, we still can solve the derivative to find $R_s^*$ numerically.

Though it seems difficult to find a general closed-form solution for~(\ref{eq:THR-OPT-RS-NAE}), we can obtain an accurate approximation in the high SNR region. Specifically, as derived in Appendix~\ref{Proof for Approximatin of Optimal Secrecy Rate RS-OPT-APP-P-NAE}, when $P\to\infty$, we have
\begin{align}
R_s^*\approx\frac{1}{N\ln\left(2\right)}\left({\rm W}_0\left(\frac{{\rm exp}\left(1\right)N!\,P^N}{\left(\sqrt{\lambda\left(\epsilon,N\right)}+1\right)^{2N}}\right)-1\right)\;,\label{eq:RS-OPT-APP-P-NAE}
\end{align}
where ${\rm W}_0\left(\cdot\right)$ is the principle branch of the Lambert-W function and $\lambda\left(\epsilon,N\right)$ is defined in~(\ref{eq:lambda}).

From~(\ref{eq:RS-OPT-APP-P-NAE}) and~\cite[Eq. 65]{Corless1997}, we know that $R_s^*$ increases with $P$. Therefore, from~(\ref{eq:OPT-PHI-MAX-PTX-NAE}) and~(\ref{eq:OPT-RB-MAX-PTX-NAE}), we can observe that for the throughput-maximizing scheme, as $P\to\infty$, the optimal power allocation ratio $\Phi^*$ and the added rate redundancy \mbox{$R_e^*=R_b^*-R_s^*$} converge to
\begin{align}
\lim_{P\to\infty}{\Phi^*}&=\frac{1}{\sqrt{\lambda\left(\epsilon,N\right)}+1}\;,\label{eq:PHI-LMT-THR-MAX-NAE}\\
\lim_{P\to\infty}{R_e^*}&=\log_2\left(\sqrt{\lambda\left(\epsilon,N\right)}+1\right)\;.\label{eq:RE-LMT-THR-MAX-NAE}
\end{align}

Furthermore, based on~(\ref{eq:RS-OPT-APP-P-NAE}), a high SNR throughput approximation for the NAE scheme is derived in Appendix~\ref{Derivation for High SNR Throughput Approximation in  Thr_Asy_APP_P_NAE}, and is given as follows:
\begin{align}
\eta_{\rm NAE}^*\left(\epsilon\right)\approx\eta_{\rm NAE}^{\epsilon=1}-\eta_{\rm NAE}^{\rm loss}\left(\epsilon\right)\;,\label{eq:Thr_Asy_APP_P_NAE}
\end{align}
where
\begin{align}
\eta_{\rm NAE}^{\epsilon=1}&\!=\!\log_2\!\left(P\right)\!-\!\frac{1}{N}\!\log_2\!\left({\ln\!\left(P\right)}\right)\!+\!\frac{1}{N}{\log_2\!\left({\!\left(N\!\!-\!\!1\right)!}\right)},\label{eq:Thr_NS_NAE}\\
\eta_{\rm NAE}^{\rm loss}\!\left(\epsilon\right)&\!=\!2\log_2\!\left(\!\sqrt{\lambda\!\left(\epsilon,\!N\right)}\!+\!1\!\right)\!.\label{eq:Thr_Loss_NAE}
\end{align}
Here $\eta_{\rm NAE}^{\epsilon=1}$ is the high SNR throughput approximation for the NAE scheme when the system is operating without any secrecy constraint (i.e., $\epsilon=1$), while $\eta_{\rm NAE}^{\rm loss}\left(\epsilon\right)$ reflects the throughput loss for realizing secure transmission. Interestingly, the throughput loss in~(\ref{eq:Thr_Loss_NAE}) is independent of $P$ and this is a direct consequence of the zero-noise assumption of Eve.

From~(\ref{eq:Thr_Asy_APP_P_NAE})-(\ref{eq:Thr_Loss_NAE}), several observations can be made:
\begin{itemize}
\item The benefits of adding extra transmit antennas are two-fold: i) the system is more capable of achieving a larger throughput, ii) the throughput loss for securing the transmission will be reduced. However, the benefits in terms of reducing $\eta_{\rm NAE}^{\rm loss}\left(\epsilon\right)$ by adding extra antennas are limited, i.e.,
    \begin{align}
    \lim_{N\to\infty}\eta_{\rm NAE}^{\rm loss}\left(\epsilon\right)=2\log_2\left(\sqrt{\ln\left(\frac{1}{\epsilon}\right)}+1\right)\;,\nonumber
    \end{align}
    which depends only on $\epsilon$, as we may expect.
\item Even in the high SNR region, the second term in~(\ref{eq:Thr_NS_NAE}) is usually insignificant due to the double logarithm. Therefore, to achieve a specified target throughput, the additional power cost for imposing or strengthening the secrecy constraint can be approximated as
    \begin{align}
    &\Delta_{P}^{\rm NAE}\left(\epsilon_1,\epsilon_2,N\right)\;{\rm (dB)}\label{eq:PWR-PNY-NAE}\\
    \approx&\begin{cases}
    20\log_{10}\left(\sqrt{\lambda\left(\epsilon_2,N\right)}+1\right)&{\rm if}\;\epsilon_2\leq\epsilon_1=1\;,\\
    \frac{10}{N-1}\log_{10}\left(\frac{\epsilon_1}{\epsilon_2}\right)&{\rm if}\;\epsilon_2\leq\epsilon_1\ll1\;,
    \end{cases}
    \nonumber
    \end{align}
    which is independent of the targeted throughput. It is clear that the power cost in both cases in~(\ref{eq:PWR-PNY-NAE}) can be effectively reduced by adding extra transmit antennas.
\end{itemize}

\begin{figure}[htbp]
\centering
\includegraphics[width=0.9\linewidth]{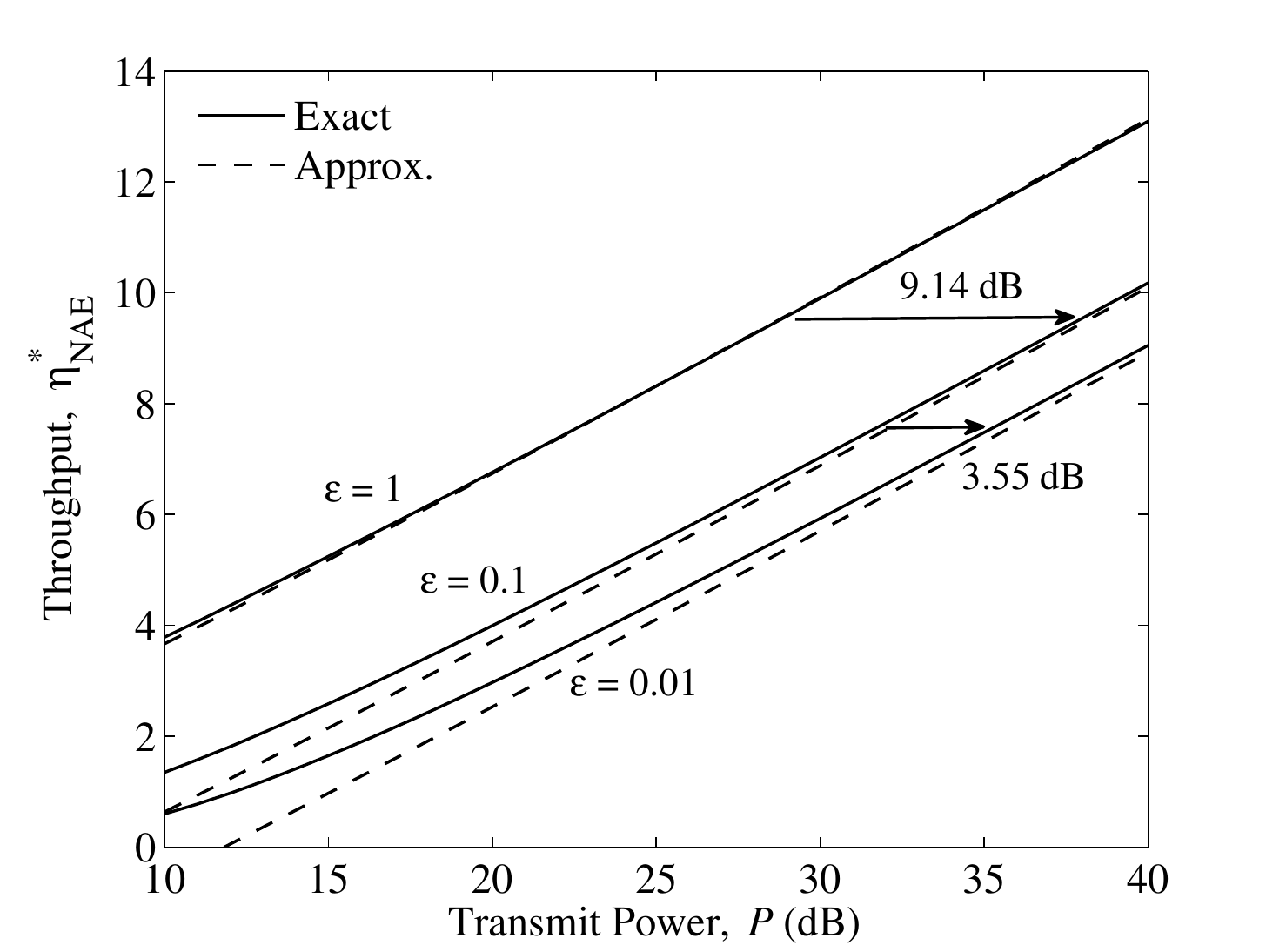}
\caption{Throughput of the NAE scheme and the high SNR approximation versus the transmit power for different secrecy constraints, with $N=4$.} \label{fig:Pwr_Pny_Epi_NAE}
\end{figure}

In Fig.~\ref{fig:Pwr_Pny_Epi_NAE}, we show that the throughput approximation in~(\ref{eq:Thr_Asy_APP_P_NAE}) is quite accurate by comparing with the exact maximal throughput, which is obtained by numerically optimizing~(\ref{eq:THR-OPT-RS-NAE}). When the secrecy constraint is strengthened, in order to maintain a specified target throughput, for $N=4$, the first arrow shows that with $\epsilon_1=1$ and $\epsilon_2=0.1$, the additional power cost is $9.14$ dB, while the second arrow suggests that with $\epsilon_1=0.1$ and $\epsilon_2=0.01$, the additional power cost is $3.55$ dB. These two values are very close to the approximation provided in~(\ref{eq:PWR-PNY-NAE}). Note that when $N$ is very small (e.g., $N=2$), the approximation in~(\ref{eq:Thr_Asy_APP_P_NAE}) becomes inaccurate due to the approximation procedure used in Appendix~\ref{Derivation for High SNR Throughput Approximation in  Thr_Asy_APP_P_NAE} and a better approximation can be obtained by plugging~(\ref{eq:RS-OPT-APP-P-NAE}) into~(\ref{eq:Thr_Max_NAE}).

\section{Secure Transmission Design\\with Adaptive Encoder}\label{Secure Transmission Design with Adaptive Encoder}
In this section, we consider the scenario where Alice dynamically adjusts the system parameters: $R_b$, $R_s$ and $\Phi$ for each realization of the channel to Bob ${\bf h}$. In other words, Alice performs adaptive encoding (AE). As before, the target is to maximize the throughput under the secrecy constraint given by a maximum allowable secrecy outage probability $\epsilon$. In particular, the values of $R_b$, $R_s$ and $\Phi$ are dynamically chosen to meet the requirement on the secrecy outage probability for each transmission, i.e., $p_{\rm so}\left(R_b\left({\bf h}\right),R_s\left({\bf h}\right),\Phi\left({\bf h}\right)\right)\leq\epsilon$. This design will in turn satisfy the requirement on the overall secrecy outage probability. With the secrecy constraint satisfied, the problem of maximizing the throughput is equivalent to maximizing $R_s\left({\bf h}\right)$ for each ${\bf h}$. Intuitively, the AE scheme can achieve a larger throughput compared with the NAE scheme, but demands a higher complexity.

\subsection{Message Rate Maximization}
From~(\ref{eq:INS-SOPH}), the problem of maximizing $R_s\left({\bf h}\right)$ for each ${\bf h}$ is given by
\begin{align}
\mathop{\max}_{R_b\left({\bf h}\right),R_s\left({\bf h}\right),\Phi\left({\bf h}\right)}\;\;&{R_s\left({\bf h}\right)}\nonumber\\
{\rm s.t.}\;\;&p_{\rm so}\left(R_b\left({\bf h}\right),R_s\left({\bf h}\right),\Phi\left({\bf h}\right)\right)\leq\epsilon\;,\nonumber\\
&{R_b}\left({\bf h}\right)\leq\log_2\left(1+P\Phi\left({\bf h}\right)||{\bf h}||^2\right)\;,\nonumber\\
&0\leq{R_s\left({\bf h}\right)}\leq{R_b\left({\bf h}\right)}\;,\nonumber\\
&0\leq\Phi\left({\bf h}\right)\leq1\;,\label{eq:RS-OPT-SEC-AE}
\end{align}
where the second constraint eliminates the capacity outages at the intended receiver.

The solution to (\ref{eq:RS-OPT-SEC-AE}) is presented as follows, whilst the derivation is relegated to Appendix~\ref{Proof for LBL-MAX-RS-AE}.

\begin{MAX-RS-AE}\label{LBL-MAX-RS-AE}
With an adaptive encoder, for a given realization of the intended channel ${\bf h}$ and under the secrecy constraint $p_{\rm so}\left(R_b\left({\bf h}\right),R_s\left({\bf h}\right),\Phi\left({\bf h}\right)\right)\leq\epsilon$, the maximal message rate in~(\ref{eq:RS-OPT-SEC-AE}) is given by:
\begin{align}
&{R_s^{\max}}\left({\bf h}\right)\label{eq:RS-MAX-AE}\\
=&2\log_2\left({\frac{\sqrt{P{\lambda\left(\epsilon,N\right)}||{\bf h}||^2}-\sqrt{P||{\bf h}||^2-{\lambda\left(\epsilon,N\right)}+1}}{{\lambda\left(\epsilon,N\right)}-1}}\right)\;,\nonumber
\end{align}
where ${\lambda\left(\epsilon,N\right)}$ is defined in~(\ref{eq:lambda}). The corresponding optimal system parameters are given by:
\begin{align}
\mu^{*}&={{\lambda\left(\epsilon,N\right)}}/{P}\;,\label{eq:OPT-MU-MAX-RS-AE}\\
\Phi^{*}\left({\bf h}\right)&=\frac{\sqrt{{\lambda\left(\epsilon,N\right)}\left(1+\frac{1-{\lambda\left(\epsilon,N\right)}}{P||{\bf h}||^2}\right)}-1}{{\lambda\left(\epsilon,N\right)}-1}\;,\label{eq:OPT-PHI-MAX-RS-AE}\\
R_b^{*}\left({\bf h}\right)&=\log_2\left(1+P\Phi^*\left({\bf h}\right)||{\bf h}||^2\right)\;.\label{eq:OPT-RB-MAX-RS-AE}
\end{align}
\end{MAX-RS-AE}

From~(\ref{eq:RS-MAX-AE})-(\ref{eq:OPT-RB-MAX-RS-AE}), several observations can be made:
\begin{itemize}
\item Intuitively, we may expect that a positive message rate can always be achieved by properly adjusting the transmission system; thus, it may not be necessary to introduce the on-off transmission protocol. However, the derivation in Appendix~\ref{Proof for LBL-MAX-RS-AE} shows that after guaranteeing the secrecy performance, a positive $R_s\left({\bf h}\right)$ can be achieved only when \mbox{$||{\bf h}||^2>{{\lambda\left(\epsilon,N\right)}}/{P}$}. In other words, when Bob's channel is not sufficiently good, a positive $R_s\left({\bf h}\right)$ and the required secrecy performance cannot be achieved simultaneously, no matter how Alice adjusts the power allocation or encoding. Therefore, to maximize the throughput whilst guaranteeing the required secrecy level, on-off transmission with $\mu={{\lambda\left(\epsilon,N\right)}}/{P}$ is introduced, as stated in~(\ref{eq:OPT-MU-MAX-RS-AE}).
\item The optimal power allocation ratio $\Phi^*\left({\bf h}\right)$ in~(\ref{eq:OPT-PHI-MAX-RS-AE}) decreases with increasing $P$, approaching the constant:
    \begin{align}
    \lim_{P\to\infty}{\Phi^*}\left({\bf h}\right)=\frac{1}{\sqrt{\lambda\left(\epsilon,N\right)}+1}\;,\label{eq:PHI-OPT-RS-MAX-AE}
    \end{align}
    which is independent of ${\bf h}$ and is identical to that for the NAE scheme in~(\ref{eq:PHI-LMT-THR-MAX-NAE}). This convergence suggests that in the high SNR region, near optimal performance can be obtained by employing a non-adaptive power allocation strategy.
\item With optimized power allocation, $R_b\left({\bf h}\right)$ is set to the instantaneous capacity of Bob's channel to guarantee successful decoding, as shown in~(\ref{eq:OPT-RB-MAX-RS-AE}). From~(\ref{eq:RS-MAX-AE}) and~(\ref{eq:OPT-RB-MAX-RS-AE}), we can observe that as $P\to\infty$, the added rate redundancy $R_e\left({\bf h}\right)=R_b\left({\bf h}\right)-R_s\left({\bf h}\right)$ converges to
    \begin{align}
    \lim_{P\to\infty}{R_e}\left({\bf h}\right)=\log_2\left(\sqrt{\lambda\left(\epsilon,N\right)}+1\right)\;,\label{eq:RE-LMT-RS-MAX-AE}
    \end{align}
    which is independent of ${\bf h}$ and is identical to that for the NAE scheme in~(\ref{eq:RE-LMT-THR-MAX-NAE}).
\item Furthermore, as shown in Appendix~\ref{Proof for LBL-MAX-RS-AE}, $R_s^{\max}\left({\bf h}\right)$ in~(\ref{eq:RS-MAX-AE}) is always achieved when $p_{\rm so}\left(R_b\left({\bf h}\right),R_s\left({\bf h}\right),\Phi\left({\bf h}\right)\right)=\epsilon$ and this implies that the overall secrecy outage probability in~(\ref{eq:AVG-SOP}) is also guaranteed to be $\bar{p}_{\rm so}=\epsilon$.
\end{itemize}

\begin{figure}[htbp]
\centering
\includegraphics[width=0.9\linewidth]{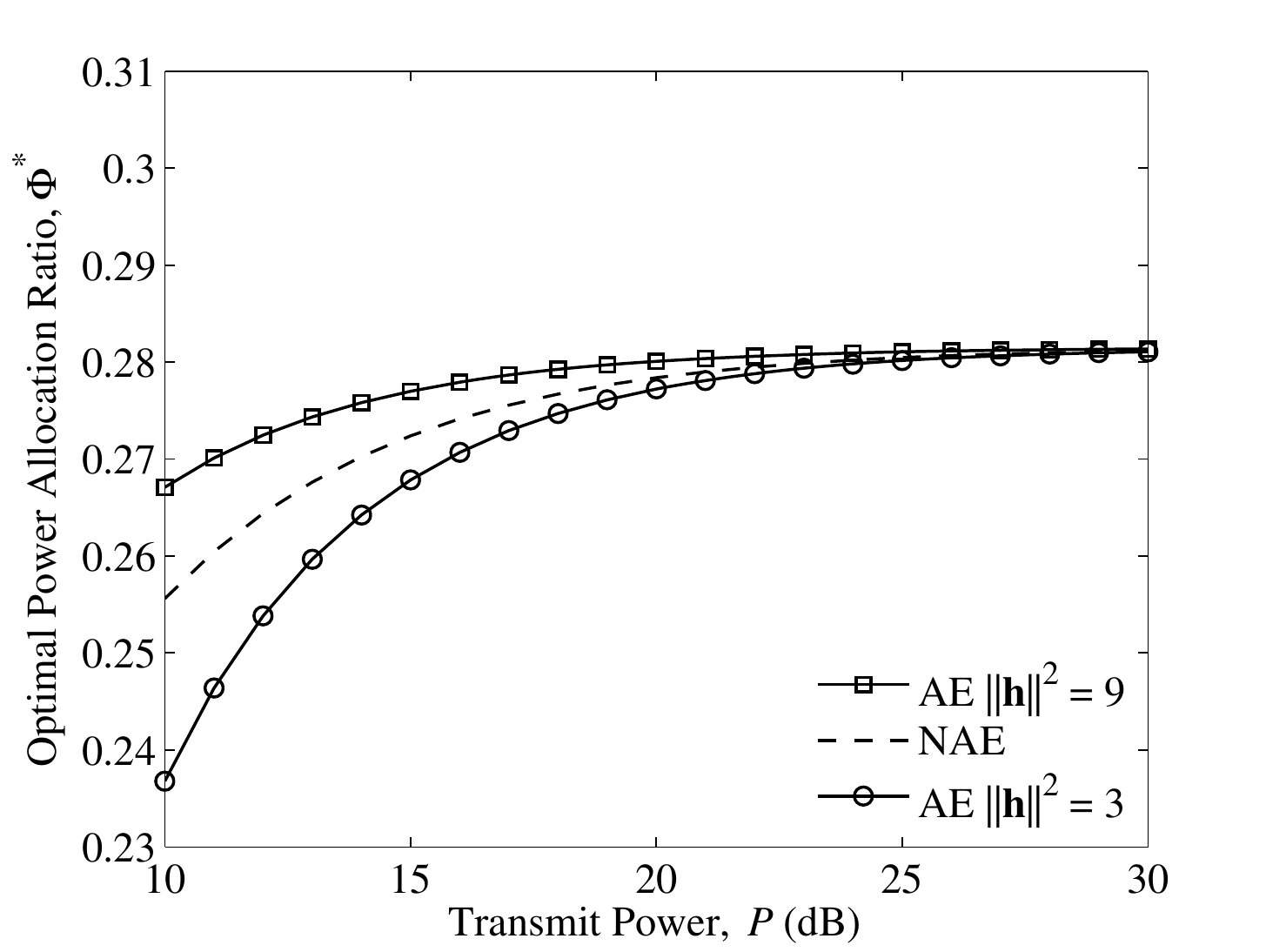}
\caption{Optimal power allocation ratios of the NAE and AE schemes versus the transmit power, with $N=8$ and $\epsilon=0.01$.}
\label{fig:Phi_Opt_P_AE_NAE}
\end{figure}

Fig.~\ref{fig:Phi_Opt_P_AE_NAE} illustrates our analysis, showing $\Phi^*\left({\bf h}\right)$ for the AE scheme for different ${\bf h}$. The value of $\Phi^*$ for the NAE scheme is also shown for comparison and it is found by numerically optimizing $R_s$ in~(\ref{eq:THR-OPT-RS-NAE}), followed by plugging it into~(\ref{eq:OPT-PHI-MAX-PTX-NAE}). As we may expect, the optimal power allocation ratios all converge to the same limit with increasing $P$. This observation is similar to the one in~\cite{Zhou2010} where the channel gain was found to be irrelevant in finding the optimal power allocation in the high SNR region for maximizing the ergodic secrecy rate in fast fading channels. However, besides the similarity, the difference is also significant. In the high SNR region, the optimal power allocation ratio that maximizes the ergodic secrecy rate for fast fading channels~\cite{Zhou2010} is around $0.5$, while the limiting value of $\Phi^*\left({\bf h}\right)$ in~(\ref{eq:PHI-OPT-RS-MAX-AE}) that maximizes the throughput for slow fading channels depends strongly on the required security level $\epsilon$ and the number of transmit antennas $N$.

\subsection{Throughput Performance Analysis}\label{Throughput Performance Analysis}
By~(\ref{eq:THR-DEF}) and~(\ref{eq:RS-MAX-AE}), the throughput for the optimized AE scheme is given by
\begin{align}
\eta_{\rm AE}^*&=\mathbb{E}_{\bf h}\left[R_s^{\max}\left({\bf h}\right)\right]\;.\label{eq:Thr_Int_AE}
\end{align}

Though it seems difficult to find a closed-form expression, similar to the NAE case, progress can be made by appealing to the high SNR region.  Specifically, as $P\to\infty$, the following approximation is derived in Appendix~\ref{Proof for High SNR Approximation of Throughput Thr_Asy_P_AE}:
\begin{align}
\eta_{\rm AE}^*&\approx\log_2\!\left(\!\frac{P}{\lambda\!\left(\epsilon,\!N\right)}\!\right)+\frac{\psi\!\left(N\right)}{\ln\!\left(2\right)}\label{eq:Thr_Asy_P_AE_Comp_AE}\\
&+2\log_2\!\left(\!\frac{\sqrt{\lambda\!\left(\epsilon,\!N\right)}}{\sqrt{\lambda\!\left(\epsilon,\!N\right)}\!+\!1}\!\right)\Tilde{\Gamma}\!\left(N,\!\frac{\lambda\!\left(\epsilon,\!N\right)}{P}\!\right)\nonumber\\
&+\frac{\left(\lambda\!\left(\epsilon,N\right)\!/P\right)^N}{N^2\!\left(N-1\right)!\ln\!\left(2\right)\!}\,{_2F_2\!\left(\!N,\!N;\!N\!+\!1,\!N\!+\!1;\!-\frac{\lambda\!\left(\epsilon,\!N\right)}{P}\!\right)\!}\;,\nonumber
\end{align}
where $\lambda\left(\epsilon,N\right)$ is defined in~(\ref{eq:lambda}), $\psi\left(N\right)$ is the digamma function, and $_2F_2\left(\cdot,\cdot;\cdot,\cdot;\cdot\right)$ is the generalized hypergeometric function~\cite[Eq. 9.14.1]{Gradshteyn2007}.

To gain more insights, as shown in Appendix~\ref{Proof for High SNR Approximation of Throughput Thr_Asy_P_AE}, when \mbox{$P\to\infty$},~(\ref{eq:Thr_Asy_P_AE_Comp_AE}) is further approximated as
\begin{align}
\eta_{\rm AE}^*\left(\epsilon\right)\approx\eta_{\rm AE}^{\epsilon=1}-\eta_{\rm AE}^{\rm loss}\left(\epsilon\right)\;,\label{eq:Thr_Asy_P_AE}
\end{align}
where
\begin{align}
\eta_{\rm AE}^{\epsilon=1}&=\log_2\left({P}\right)+\frac{\psi\left(N\right)}{\ln(2)}\;,\label{eq:Thr_NS_AE}\\
\eta_{\rm AE}^{\rm loss}\left(\epsilon\right)&=2\log_2\left(\sqrt{\lambda\left(\epsilon,N\right)}+1\right)\;.\label{eq:Thr_Loss_AE}
\end{align}
Here $\eta_{\rm AE}^{\epsilon=1}$ is the high SNR throughput approximation for the AE scheme without any secrecy constraint, which coincides with the high SNR approximation of the ergodic capacity of the MISO Rayleigh fading channel~\cite{Tse2005}, while $\eta_{\rm AE}^{\rm loss}\left(\epsilon\right)$ reflects the throughput loss incurred by enforcing the secrecy constraint.

From~(\ref{eq:Thr_Asy_P_AE})-(\ref{eq:Thr_Loss_AE}), several observations can be made:
\begin{itemize}
\item In the high SNR region, the throughput loss for imposing the secrecy constraint in~(\ref{eq:Thr_Loss_AE}) is independent of $P$ and it is identical to the throughput loss for the NAE scheme in~(\ref{eq:Thr_Loss_NAE}). The throughput loss in~(\ref{eq:Thr_Loss_AE}) is actually twice the limiting value of the rate redundancy in~(\ref{eq:RE-LMT-RS-MAX-AE}). This can be explained by noting that in the high SNR region, the transmit threshold in~(\ref{eq:OPT-MU-MAX-RS-AE}) is close to zero and, in addition to the rate redundancy in~(\ref{eq:RE-LMT-RS-MAX-AE}), the transmission system also suffers a power loss since a certain fraction of power is used to generate artificial noise, as shown in~(\ref{eq:PHI-OPT-RS-MAX-AE}). It is interesting that this power loss leads to a data rate loss which is identical with the rate redundancy in~(\ref{eq:RE-LMT-RS-MAX-AE}).
\item In the high SNR region, to achieve a specified target throughput, the additional power cost for imposing or strengthening the secrecy constraint can be approximated from~(\ref{eq:Thr_Asy_P_AE}) and the results are identical to those for the NAE scheme in~(\ref{eq:PWR-PNY-NAE}). However, although the NAE and AE schemes suffer the same power cost when imposing the same secrecy constraint, the AE scheme still outperforms the NAE scheme in terms of the throughput performance. This is due to the fact that when the secrecy constraint is removed, the AE scheme is designed to achieve a better throughput performance, compared with the NAE scheme. Specifically, when $\epsilon=1$, for both schemes, all the transmit power is allocated to the information-bearing signal and the data will be transmitted without wiretap coding. In this case, the NAE scheme still chooses a constant rate to transmit and the transmit threshold is set to the minimum required effective channel gain to support the selected rate; while the AE scheme always transmits at the instantaneous capacity of Bob's channel and the transmit threshold is therefore set to zero.
\end{itemize}

\begin{figure}[htbp]
\centering
\includegraphics[width=0.9\linewidth]{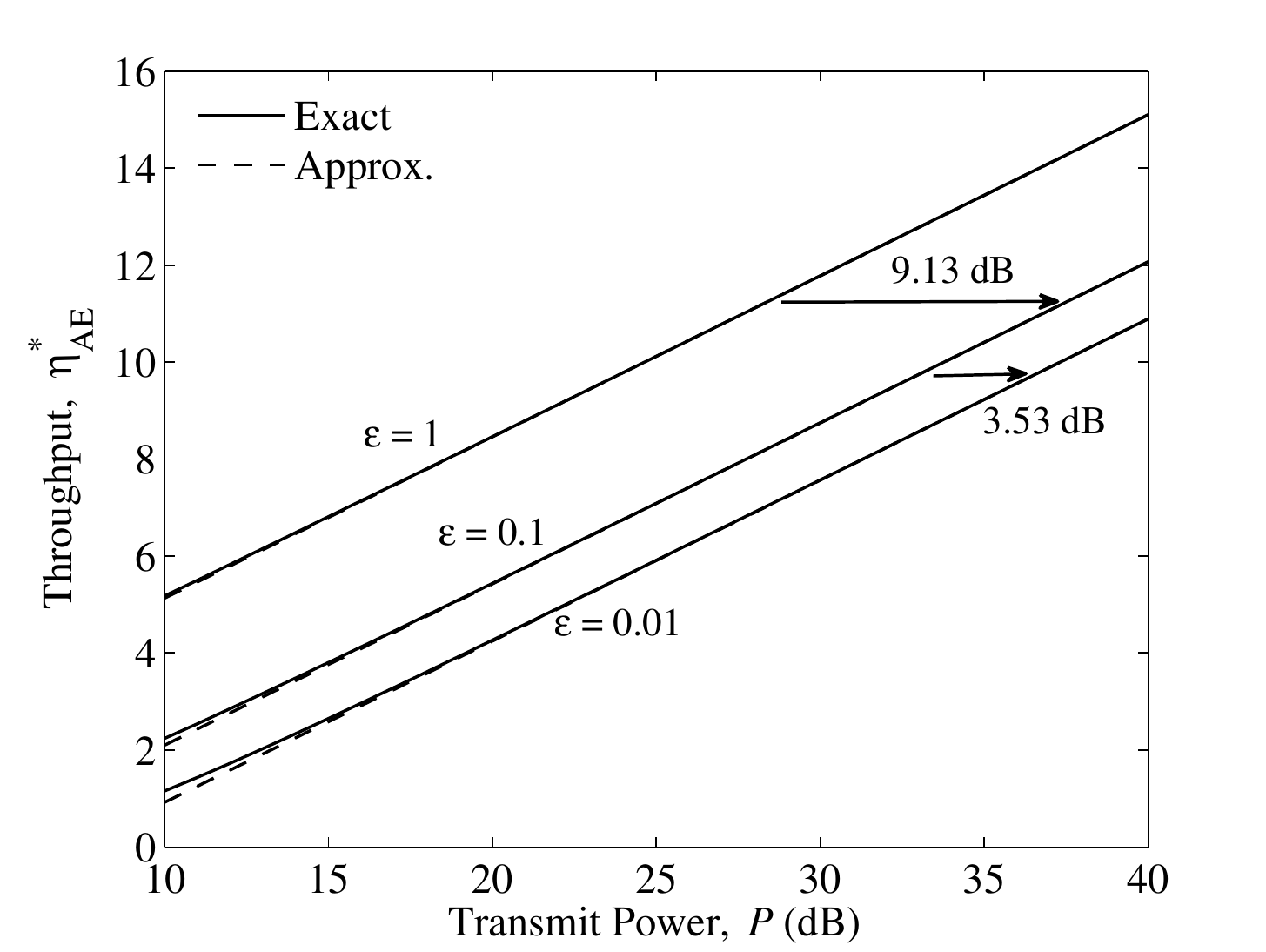}
\caption{Throughput of the AE scheme and the high SNR approximation versus the transmit power for different secrecy constraints, with $N=4$.}
\label{fig:Pwr_Pny_Epi_AE}
\end{figure}

In Fig.~\ref{fig:Pwr_Pny_Epi_AE}, we show that the throughput approximation in~(\ref{eq:Thr_Asy_P_AE}) is quite accurate by comparing with the exact maximal throughput, which is obtained by numerically evaluating~(\ref{eq:Thr_Int_AE}). The additional power cost analysis is confirmed by the examples therein. Note that when $N$ is very small (e.g., $N=2$),~(\ref{eq:Thr_Asy_P_AE}) becomes less accurate due to the approximation procedure used in Appendix~\ref{Proof for High SNR Approximation of Throughput Thr_Asy_P_AE} and~(\ref{eq:Thr_Asy_P_AE_Comp_AE}) provides a better approximation.

\subsection{Throughput Gain of Adaptation: AE over NAE}
In the following, we analyze the throughput gain brought by doing adaptation. With the throughput approximations for the NAE and AE schemes in~(\ref{eq:Thr_Asy_APP_P_NAE}) and~(\ref{eq:Thr_Asy_P_AE}), in the high SNR region, the throughput gain can be approximated by
\begin{align}
&{\eta_{\rm AE}^*}-{\eta_{\rm NAE}^*}\label{eq:Thr_Gain}\\
\approx&\frac{1}{N}\log_2\left({\ln\left(P\right)}\right)+\left(\frac{\psi\left(N\right)}{\ln\left(2\right)}-{\frac{1}{N}\log_2\left(\left(N-1\right)!\right)}\right)\;.\nonumber
\end{align}
With this approximation, several observations can be made:
\begin{itemize}
\item In the high SNR region, the throughput gain increases with $P$, albeit very slowly, as can be seen from the double logarithm in the first term of~(\ref{eq:Thr_Gain}).
\item It turns out that as $N$ increases, the second term in~(\ref{eq:Thr_Gain}) (i.e., the $P$-independent term) increases slowly. The first term, on the other hand, decreases very fast with $N$ (i.e., linearly), and moreover, this term becomes dominant when $P$ is large. These observations imply that for high SNR, the throughput gap between the NAE and AE schemes shrinks when the number of antennas grows. Consequently, for systems operating at high SNRs, if the number of antennas is not small, it may be more preferable to employ the NAE scheme rather than the AE scheme, due to the complexity saving.
\item The throughput gain approximation in~(\ref{eq:Thr_Gain}) is independent of the secrecy constraint $\epsilon$. This can be explained by noting that in the high SNR region, the NAE and AE schemes incur the same throughput loss for imposing the secrecy constraint, as shown in~(\ref{eq:Thr_Asy_APP_P_NAE}) and~(\ref{eq:Thr_Asy_P_AE}).
\end{itemize}

\begin{figure}[htbp]
\centering
\includegraphics[width=0.9\linewidth]{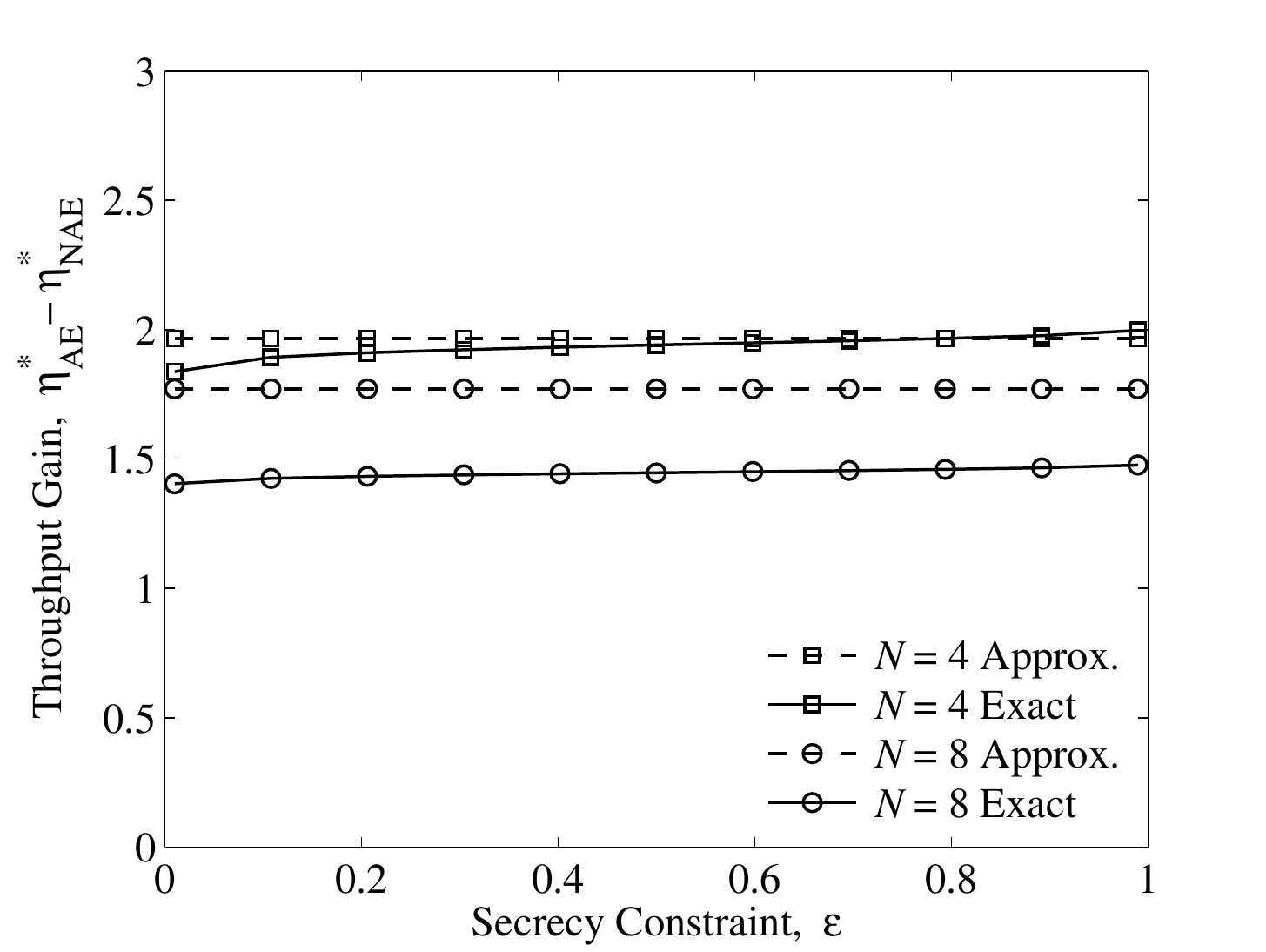}
\caption{Throughput gain of the AE scheme over the NAE scheme and the high SNR approximation versus the secrecy constraint for different number of transmit antennas, with $P=40$ dB.}
\label{fig:Thr_Cmp_P_EPI_NAE_AE}
\end{figure}

Fig.~\ref{fig:Thr_Cmp_P_EPI_NAE_AE} compares the approximation in~(\ref{eq:Thr_Gain}) with the exact throughput gain. It is clear that the exact throughput gain decreases with increasing $N$. Note that the variation in the throughput gain is not very significant over a wide range of $\epsilon$, though the gain does decrease with strengthening  the secrecy constraint (i.e., reducing $\epsilon$). We note that the approximation is not very accurate, since the throughput gain is relatively small and thus the approximation error in~(\ref{eq:Thr_Gain}) is relatively pronounced. What is important is that, as stated above, our analysis accurately predicts the varying trends of the exact throughput gain.

\section{Conclusion}
This paper investigated the design of artificial-noise-aided secure multi-antenna transmission in slow fading channels. We provided explicit design solutions to the secrecy-constrained throughput-maximization problem with either non-adaptive transmission (i.e., the NAE scheme) or adaptive transmission (i.e., the AE scheme). To facilitate practical system design, we examined the additional power cost for achieving a higher secrecy level for both schemes and analyzed the throughput gain of the AE scheme over the NAE scheme in the high SNR region. Our analysis obtained new insights on the design of artificial-noise-aided secure multi-antenna transmission in slow fading channels.

\appendix
\subsection{Proposition~\ref{LBL-MAX-PTX-NAE}}\label{Proof for LBL-MAX-PTX-NAE}
In this proof, we derive the optimal values of the system parameters $R_b$, $\mu$ and $\Phi$ to maximize the transmit probability $p_{\rm tx}$  in~(\ref{eq:PTX-OPT-SEC-NAE}) (i.e., minimize the average delay). We first optimize $R_b$ and $\mu$ for a given $\Phi$, and then optimize $\Phi$. The message rate $R_s$ is assumed to be at a prescribed value.

As we discussed at the beginning of Section~\ref{Secure Transmission Design with Non-Adaptive Encoder}, for the NAE scheme, the overall secrecy outage probability in~(\ref{eq:AVG-SOP}) reduces to~(\ref{eq:INS-SOPH}). For a given $\Phi$, from~(\ref{eq:INS-SOPH}), it is clear that the secrecy constraint in~(\ref{eq:PTX-OPT-SEC-NAE}) can be satisfied by choosing a large enough $R_b$. Substituting~(\ref{eq:INS-SOPH}) into the secrecy constraint in~(\ref{eq:PTX-OPT-SEC-NAE}) (i.e., $\bar{p}_{\rm so}\left(R_b,\Phi\right)\leq\epsilon$) and solving w.r.t. $R_b$, we have
\begin{align}
R_b\geq{R_s}+\log_2\left(1+\frac{\lambda\Phi}{1-\Phi}\right)=:{R_b^{\min}}\;,\nonumber
\end{align}
where $\lambda$ is defined in~(\ref{eq:lambda}) and its parameters are omitted for brevity.

To minimize the average delay, the transmit probability $p_{\rm tx}$ in~(\ref{eq:PTX-NAE}) needs to be maximized. From~(\ref{eq:PTX-NAE}), it is clear that $p_{\rm tx}$ is maximized when the transmit threshold $\mu$ is set to its smallest possible value. From the second constraint in~(\ref{eq:PTX-OPT-SEC-NAE}), which eliminates capacity outages at Bob, we know that for a given $\Phi$, $\mu$ should be set to
\begin{align}
\mu_{\min}:=\frac{1}{P\Phi}{\left(2^{R_b^{\min}}-1\right)}\;,\nonumber
\end{align}
in order to maximize the transmit probability, whilst satisfying the secrecy constraint.

Substituting $\mu_{\min}$ into~(\ref{eq:PTX-NAE}), the corresponding $p_{\rm tx}$ is given by
\begin{align}
p_{\rm tx}\left(\Phi\right)=\Tilde{\Gamma}\left(N,\omega\left(\Phi\right)\right)\;,\label{eq:PTX-PHI-NAE}
\end{align}
where
\begin{align}
\omega\left(\Phi\right)=\frac{\rho}{1-\Phi}+\frac{\varrho}{\Phi},\;\;\rho=\frac{2^{R_s}\lambda}{P},\;\;\varrho=\frac{2^{R_s}-1}{P}\;.
\end{align}

Since the regularized upper incomplete gamma function decreases with its second parameter, it is clear that~$p_{\rm tx}\left(\Phi\right)$ in~(\ref{eq:PTX-PHI-NAE}) achieves its maximum when $\omega\left(\Phi\right)$ takes its minimum. Note that $\omega\left(\Phi\right)$ is a convex function of~$\Phi$; thus, by solving the derivative of $\omega\left(\Phi\right)$ w.r.t. $\Phi$, the optimal power allocation ratio can be found as
\begin{align}
\Phi^{*}&=\frac{\sqrt{\varrho}}{\sqrt{\rho}+\sqrt{\varrho}}\;,\nonumber
\end{align}
and the corresponding maximal transmit probability $p_{\rm tx}^{\max}$ is given by
\begin{align}
p_{\rm tx}^{\max}:=p_{\rm tx}\left(\Phi^*\right)=\Tilde{\Gamma}\left(N,\left(\sqrt{\rho}+\sqrt{\varrho}\right)^2\right)\;.\nonumber
\end{align}
Summarizing the obtained results, we have Proposition~\ref{LBL-MAX-PTX-NAE}.

\subsection{Monotonicity of Quantity in~(\ref{eq:lambda})}\label{Proof for Monotonicity of Quantity lambda}
In this proof, we show that $\lambda$ in~(\ref{eq:lambda}) decreases with increasing $N$. To this end, we temporarily assume that $N$ is a continuous variable. Taking the derivative of $\lambda$ w.r.t. $N$, we have
\begin{align}
\frac{{\rm d}\lambda}{{\rm d}N}=\epsilon^\frac{1}{1-N}\left(1+\frac{\ln\left(\epsilon\right)}{N-1}\right) -1\;.\label{eq:DER-LAM-NAE}
\end{align}
To show that the above derivative is negative, we need to show that the expression in the bracket is smaller than $\epsilon^\frac{1}{N-1}$. Since the secrecy constraint $\epsilon$ is smaller than one, we know that $\frac{\ln\left(\epsilon\right)}{N-1}<0$. From the inequality $1+x<\mathrm{e}^x$ for $x<0$, letting $x=\frac{\ln\left(\epsilon\right)}{N-1}$, we see that the quantity in the bracket in~(\ref{eq:DER-LAM-NAE}) satisfies
\begin{align}
1+\frac{\ln\left(\epsilon\right)}{N-1}<\exp\left(\frac{\ln\left(\epsilon\right)}{N-1}\right)=\epsilon^\frac{1}{N-1}\;.\nonumber
\end{align}
Thus, we know that the derivative in~(\ref{eq:DER-LAM-NAE}) is negative. Since the discrete $N$ is just a sampled version of the continuous $N$, the monotonic properties are the same. We then know that $\lambda$ in~(\ref{eq:lambda}) decreases with increasing $N$.

\subsection{Approximation of Optimal Message Rate in~(\ref{eq:RS-OPT-APP-P-NAE})}\label{Proof for Approximatin of Optimal Secrecy Rate RS-OPT-APP-P-NAE}
In this proof, we derive an approximation for the optimal message rate $R_s^*$ in~(\ref{eq:THR-OPT-RS-NAE}). From the discussions after~(\ref{eq:Thr_Max_NAE}), we know that one can solve the derivative of the throughput $\eta_{\rm NAE}$ w.r.t. $R_s$ to find $R_s^*$. However, the equality obtained by setting the derivative to zero is quite complicated and seems not solvable. Hence, we consider first approximating the throughput and then optimizing the approximated throughput. The details can be found as follows.

Using the inequality $2^{R_s}>2^{R_s}-1$, a lower bound to $\eta_{\rm NAE}$ can be given by
\begin{align}
\eta_{\rm NAE}\geq{\log_2\left(x\right)}\times\Tilde{\Gamma}\left(N,{\varsigma}x\right)=:\eta_{\rm NAE}^{\rm LB}\;,\label{eq:LB-THR-NAE}
\end{align}
where $x=2^{R_s}$, ${\varsigma}=\frac{1}{P}\left(\sqrt{\lambda}+1\right)^2$ and $\lambda$ is defined in~(\ref{eq:lambda}).

As $P\to\infty$, it is clear that ${\varsigma}\to0$. Using the series representation of the regularized upper incomplete gamma function~\cite[Eq. 8.352.4.*]{Gradshteyn2007}, we expand $\eta_{\rm NAE}^{\rm LB}$ in~(\ref{eq:LB-THR-NAE}) as
\begin{align}
\eta_{\rm NAE}^{\rm LB}&={\log_2\left(x\right)}\,{\rm e}^{-\varsigma{x}}\sum_{k=0}^{N-1}\frac{\left(\varsigma{x}\right)^k}{k!}\nonumber\\
&={\log_2\left(x\right)}\,{\rm e}^{-\varsigma{x}}\left({\rm e}^{\varsigma{x}}-\sum_{k=N}^{\infty}\frac{\left(\varsigma{x}\right)^k}{k!}\right)\nonumber\\
&={\log_2\left(x\right)}\left(1-\sum_{l=0}^{\infty}\frac{\left(-\varsigma{x}\right)^l}{l!}\sum_{k=N}^{\infty}\frac{\left(\varsigma{x}\right)^k}{k!}\right)\label{eq:SER-REP-EXP-NAE}\\
&={\log_2\left(x\right)}\left(1-\frac{\left(\varsigma{x}\right)^N}{N!}+\mathcal{O}\left(\varsigma^{N+1}\right)\right)\;,\label{eq:Thr_LB_SE_NAE}
\end{align}
where in~(\ref{eq:SER-REP-EXP-NAE}) we used the standard series expansion of the exponential function.

Ignoring the high order terms in the above series representation and taking the derivative w.r.t. $x$, we have
\begin{align}
\frac{{\rm d}\tilde{\eta}_{\rm NAE}^{\rm LB}}{{\rm d}x}=\frac{\varsigma^N}{x\ln\left(2\right)N!}\left(N!\varsigma^{-N}-x^N\left(1+\ln\left(x^N\right)\right)\right)\;.\nonumber
\end{align}
We then set the above derivative to zero and solve the obtained equality w.r.t. $x$. After summarizing the obtained results, we have the approximation in~(\ref{eq:RS-OPT-APP-P-NAE}).

\subsection{High SNR Throughput Approximation in~(\ref{eq:Thr_Asy_APP_P_NAE})}\label{Derivation for High SNR Throughput Approximation in  Thr_Asy_APP_P_NAE}
In~(\ref{eq:RS-OPT-APP-P-NAE}), we have obtained an approximation for the optimal message rate $R_s^*$ in~(\ref{eq:THR-OPT-RS-NAE}). Plugging an approximation for~(\ref{eq:RS-OPT-APP-P-NAE}) into a lower bound to the throughput in~(\ref{eq:Thr_LB_SE_NAE}) and ignoring the high order terms, we can get an elegant approximation for the maximal throughput $\eta_{\rm NAE}^*$. The details can be found as follows.

In~\cite[Eq. 65]{Corless1997}, a series expansion is provided for the principle branch of the Lambert-W function ${\rm W}_0\left(\cdot\right)$, and it states that as $x\to\infty$,
\begin{align}
{\rm W}_0\left(x\right)=\ln\left(\frac{x}{\ln\left(x\right)}\right)+\mathcal{O}\left(\frac{\ln\left(\ln\left(x\right)\right)}{\ln\left(x\right)}\right)\;.\nonumber
\end{align}
Thus, as $P\to\infty$, by ignoring the high order terms, (\ref{eq:RS-OPT-APP-P-NAE}) can be further approximated as
\begin{align}
\tilde{R}_s^*=\frac{1}{N}\log_2\left(\frac{\frac{N!P^N}{\left(\sqrt{\lambda}+1\right)^{2N}}}{\ln\left(\frac{{\rm exp}\left(1\right)N!P^N}{\left(\sqrt{\lambda}+1\right)^{2N}}\right)}\right)\;,\nonumber
\end{align}
where $\lambda$ is defined in~(\ref{eq:lambda}). Note that as $P\to\infty$, \mbox{$\frac{2^{\tilde{R}_s^*}}{P}\to0$}. Substituting the above approximation into the throughput lower bound in~(\ref{eq:Thr_LB_SE_NAE}), we have
\begin{align}
\eta_{\rm NAE}^*\approx{\tilde{R}}_s^*\left(1-\mathcal{O}\left(\left(\frac{2^{\tilde{R}_s^*}}{P}\right)^N\right)\right)\;.\nonumber
\end{align}
Taking the leading order term yields
\begin{align}
{\eta_{\rm NAE}^*}&\approx\log_2\left(P\right)-\frac{1}{N}\log_2\left({\ln\left(P\right)}\right)+\frac{1}{N}{\log_2\left({N!}\right)}\nonumber\\
&-\frac{1}{N}\log_2\left({N+\mathcal{O}\left(\frac{1}{\ln\left(P\right)}\right)}\right)-2\log_2\left(\sqrt{\lambda}+1\right)\;.\nonumber
\end{align}
Here $\mathcal{O}\left(\frac{1}{\ln\left(P\right)}\right)=\frac{1}{\ln\left(P\right)}\left(1+\ln\left({N!}\right)-2N\ln\left(\sqrt{\lambda}+1\right)\right)$, which can be ignored for high $P$, thereby giving~(\ref{eq:Thr_Asy_APP_P_NAE}).

\subsection{Proposition~\ref{LBL-MAX-RS-AE}}\label{Proof for LBL-MAX-RS-AE}
In this proof, we derive the optimal values of the system parameters $R_b$, $R_s$ and $\Phi$ to maximize the message rate $R_s$ in~(\ref{eq:RS-OPT-SEC-AE}) for a given realization of the intended channel ${\bf h}$, whilst satisfying the secrecy constraint. (Note that for the AE scheme, the design parameters $R_b$, $R_s$ and $\Phi$ are intrinsically functions of ${\bf h}$.) We first optimize $R_b$ and~$R_s$ for a given $\Phi$, and then optimize $\Phi$.

For a given $\Phi$, to eliminate capacity outages at Bob (i.e., the second constraint in~(\ref{eq:RS-OPT-SEC-AE})), Alice sets $R_b$ to the capacity of Bob's channel:
\begin{align}
R_b\left(\Phi\right)=C_b=\log_2\left(1+P\Phi||{\bf h}||^2\right)\;.\nonumber
\end{align}
For a given $\Phi$, if Alice decides to transmit with $R_b\left(\Phi\right)$ above, and with $R_s<R_b\left(\Phi\right)$, the secrecy outage probability for this transmission is given by~(\ref{eq:INS-SOPH}) as
\begin{align}
{p_{\rm so}}\!\left(R_b\!\left(\Phi\right)\!,\!R_s,\!\Phi\right)\!=\!\left(\!1\!+\!\left(2^{R_b\left(\Phi\right)-R_s}\!-\!1\right)\!\left({\frac{\Phi^{-1}\!-\!1}{N\!-\!1}}\right)\!\right)^{1-N}\;,\nonumber
\end{align}
which simply increases with increasing $R_s$. Thus, the maximal $R_s$ will be obtained when the secrecy constraint in~(\ref{eq:RS-OPT-SEC-AE}) is met with equality, i.e., $p_{\rm so}\left(R_b\left(\Phi\right),R_s,\Phi\right)=\epsilon$. Solving this equality gives
\begin{align}
R_s\left(\Phi\right)=\log_2\left(\frac{1+\tau\Phi}{1+\frac{\lambda\Phi}{1-\Phi}}\right)\;,\label{eq:RS-PHI-AE}
\end{align}
which is a function of $\Phi$, $\tau=P||{\bf h}||^2$ and $\lambda$ is defined in~(\ref{eq:lambda}).

From~(\ref{eq:RS-PHI-AE}), we see that when $\tau\leq\lambda$ (equivalently, \mbox{$||{\bf h}||^2\leq\frac{\lambda}{P}$}), it is impossible to adjust $\Phi\in\left(0,1\right)$ to achieve a positive $R_s$. For the alternative case, when $\tau>\lambda$ (i.e.,  \mbox{$||{\bf h}||^2>\frac{\lambda}{P}$}), the range of $\Phi$ which can guarantee that \mbox{$R_s\left(\Phi\right)>0$} is $(0,1-\frac{\lambda}{\tau})$. Henceforth, we focus on the case where $\tau>\lambda$.

Since $R_s\left(\Phi\right)$ in~(\ref{eq:RS-PHI-AE}) is non-concave on $\Phi$ and the logarithm function is monotonic, we then maximize an equivalent objective function $\frac{1+\tau\Phi}{1+\frac{\lambda\Phi}{1-\Phi}}$, which is concave on $\Phi\in\left(0,1\right)$ for $\tau>\lambda$. Solving the derivative of $\frac{1+\tau\Phi}{1+\frac{\lambda\Phi}{1-\Phi}}$ w.r.t. $\Phi$ leads to
\begin{align}
\Phi^*=\frac{\sqrt{\tau\lambda(\tau-\lambda+1)}-\tau}{\tau\lambda-\tau}\;,\nonumber
\end{align}
while it is also confirmed that $0<\Phi^*<1-\frac{\lambda}{\tau}$ when $\tau>\lambda$. Summarizing the obtained results, we have Proposition~\ref{LBL-MAX-RS-AE}.

\subsection{High SNR Throughput Approximation in~(\ref{eq:Thr_Asy_P_AE})}\label{Proof for High SNR Approximation of Throughput Thr_Asy_P_AE}
Having optimized the message rate for each realization of the intended channel, the average throughput can be computed by averaging the maximum message rate in~(\ref{eq:RS-MAX-AE}) over all channel realizations. Plugging~(\ref{eq:RS-MAX-AE}) into~(\ref{eq:Thr_Int_AE}), we have
\begin{align}
\eta_{\rm AE}^*&\!=\!{2}\!\!\int _{{\frac{{\lambda}}{P}}}^{\infty}\!\!\!\!\!\!\log_2\!\left(\!{\frac{\sqrt{P{\lambda}r}\!-\!\!\sqrt{Pr\!-\!\left({\lambda}\!-\!1\right)}}{{\lambda}\!-\!1}}\!\right)\!{{\rm e}^{\!-r}}\!\frac{{r}^{N\!-\!1}}{\!\left(N\!-\!1\right)!}{{\rm d}r},\label{eq:THR-INT-AE}
\end{align}
where $\lambda$ is defined in~(\ref{eq:lambda}) and the lower limits comes from the transmit threshold in~(\ref{eq:OPT-MU-MAX-RS-AE}).

It seems difficult to find a closed-form expression for the above integral; instead, we appeal to the high SNR region. The quantity ${\lambda}-1$ is independent of $P$; thus, it is negligible compared with $Pr$ when $P$ is large. Therefore, by omitting the quantity $\lambda-1$ under the second square root sign in~(\ref{eq:THR-INT-AE}), we have
\begin{align}
\eta_{\rm AE}^*&\!\approx\!2\log_2\!\!\left(\!\frac{\sqrt{P}}{\sqrt{{\lambda}}+1}\!\right)\!\Tilde{\Gamma}\!\left(\!N,\!\frac{{\lambda}}{P}\!\right)\!+\!\int_{\frac{{\lambda}}{P}}^{\infty}\!\!\!\!\!\log_2\!\left(r\right)\!{\rm e}^{-r}\!\frac{r^{N\!-\!1}}{\!\left(N\!-\!1\right)!}{\rm d}r\nonumber\\
&\!=\!2\log_2\!\!\left(\!\frac{\sqrt{{\lambda}}}{\sqrt{{\lambda}}+1}\!\right)\!\Tilde{\Gamma}\!\left(\!N,\!\frac{{\lambda}}{P}\!\right)\!+\!\frac{{\lambda}}{P}\,\frac{{\rm T}\!\left(3,\!N,\!\frac{{\lambda}}{P}\right)}{\ln\!\left(2\right)\!\left(N\!-\!1\right)!}\;,
\label{eq:Thr_Asy_P_AE_1}
\end{align}
where we used to the integral identities in~\cite[Eq. 4.358.1.6]{Gradshteyn2007} and~\cite[Eq. 29]{Geddes1990}, and ${\rm T}\left(3,N,x\right)$ is a special case for the Meijer G-function~\cite[Eq. 9.301]{Gradshteyn2007} with parameters:
\begin{align}
{\rm T}\left(3,N,x\right)={\rm G}_{2,3}^{3,0}\left(x\bigg{|}
                                      \begin{array}{c}
                                        0,0 \\
                                        -1,-1,N-1 \\
                                      \end{array}
\right)\;.\nonumber
\end{align}
By~\cite[Eq. 37 \& 38]{Geddes1990},~(\ref{eq:Thr_Asy_P_AE_1}) can be equivalently expressed as
\begin{align}
\eta_{\rm AE}^*&\approx\log_2\left(\frac{P}{{\lambda}}\right)+2\log_2\left(\frac{\sqrt{{\lambda}}}{\sqrt{{\lambda}}+1}\right)\Tilde{\Gamma}\left(N,\frac{{\lambda}}{P}\right)\nonumber\\
&+\frac{\left({\lambda}/P\right)^N}{\left(N-1\right)!\ln(2)}\sum_{k=0}^{\infty}\frac{\left(-{\lambda}/P\right)^k}{k!\left(N+k\right)^2}+\frac{\psi\left(N\right)}{\ln(2)}\;.\nonumber
\end{align}
We then note that $\frac{1}{\left(N+k\right)^2}=\frac{1}{N^2}\frac{\left(N\right)_k\left(N\right)_k}{\left(N+1\right)_k\left(N+1\right)_k}$, where \mbox{$\left(N\right)_k=\frac{\left(N+k-1\right)!}{\left(N-1\right)!}$} for $k\geq0$ is the Pochhammer symbol. Thus, the infinite summation can be expressed in terms of a hypergeometric function~\cite[Eq. 9.14.1]{Gradshteyn2007} and we have the result in~(\ref{eq:Thr_Asy_P_AE_Comp_AE}). By definition, with a large $P$, the regularized incomplete gamma function in the second term of the equation above is close to one while the third term is with order $\mathcal{O}\left(P^{-N}\right)$; thus, we have the result in~(\ref{eq:Thr_Asy_P_AE}).

\bibliographystyle{IEEEtran}
\bibliography{Cited}

\begin{IEEEbiography}[{\includegraphics[width=1in,height=1.25in,clip,keepaspectratio]{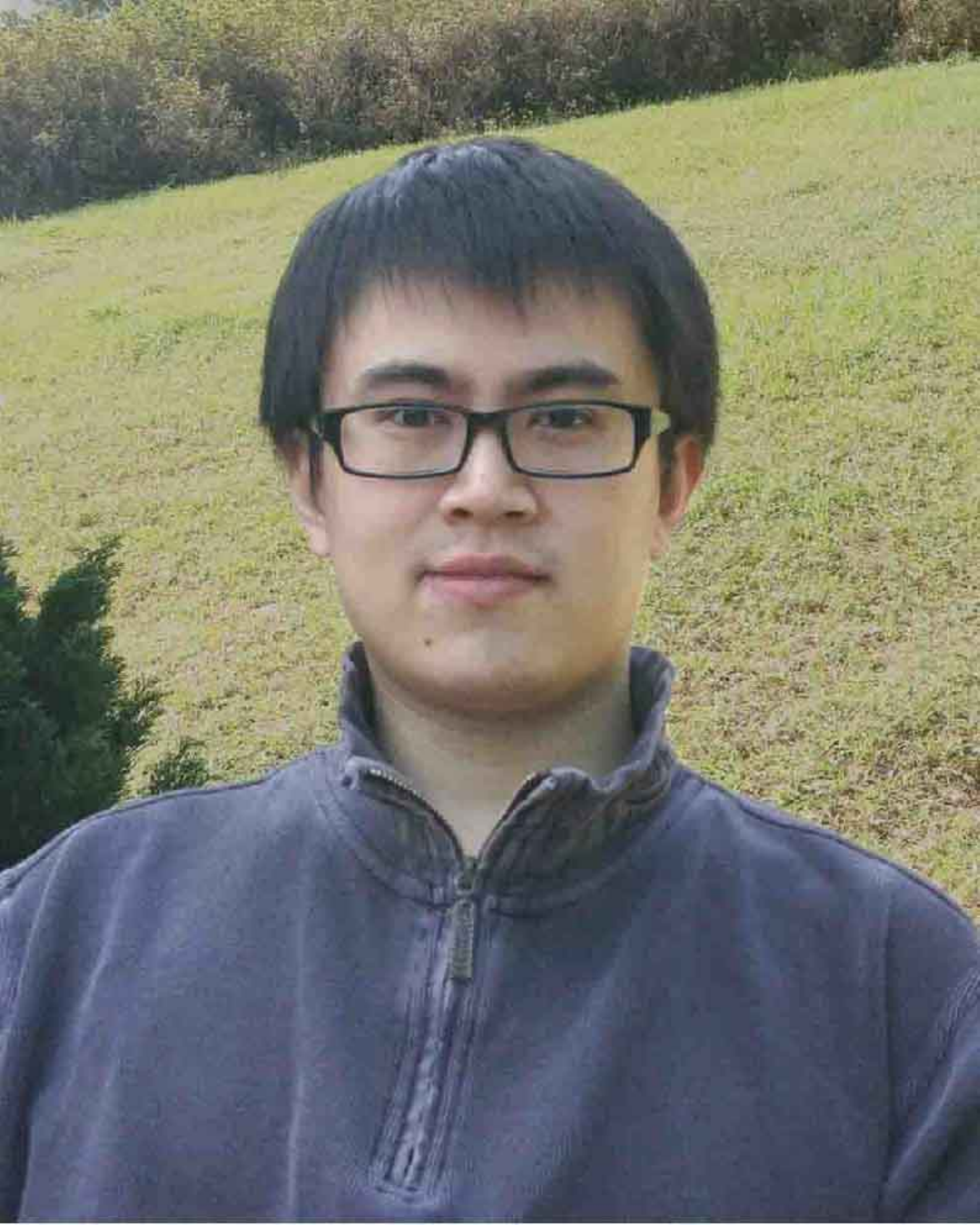}}]{Xi Zhang} (S'11) received the B.E. degree in School of Communication and Information Engineering from University of Electronic Science and Technology of China in 2010. He is currently a Ph.D. candidate at the Department of Electronic and Computer Engineering in Hong Kong University of Science and Technology. His research interests are in the fields of wireless communication and signal processing techniques, including physical-layer security, ad-hoc networking and random matrix theory.
\end{IEEEbiography}

\begin{IEEEbiography}[{\includegraphics[width=1in,height=1.25in,clip,keepaspectratio]{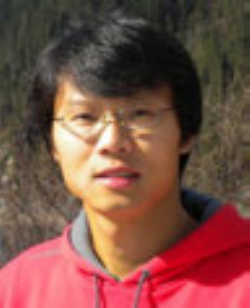}}]{Xiangyun Zhou} (S'08-M'11) is a lecturer at the Australian National University (ANU), Australia. He received the B.E. (hons.) degree in electronics and telecommunications engineering and the Ph.D. degree in telecommunications engineering from the ANU in 2007 and 2010, respectively. From June 2010 to June 2011, he worked as a postdoctoral fellow at UNIK - University Graduate Center, University of Oslo, Norway. His research interests are in the fields of communication theory and wireless networks. Dr. Zhou serves on the editorial board of Security and Communication Networks Journal (Wiley) and Ad Hoc \& Sensor Wireless Networks Journal. He has also served as the TPC member of major IEEE conferences. He is a recipient of the Best Paper Award at the 2011 IEEE International Conference on Communications.
\end{IEEEbiography}

\vfill

\begin{IEEEbiography}[{\includegraphics[width=1in,height=1.25in,clip,keepaspectratio]{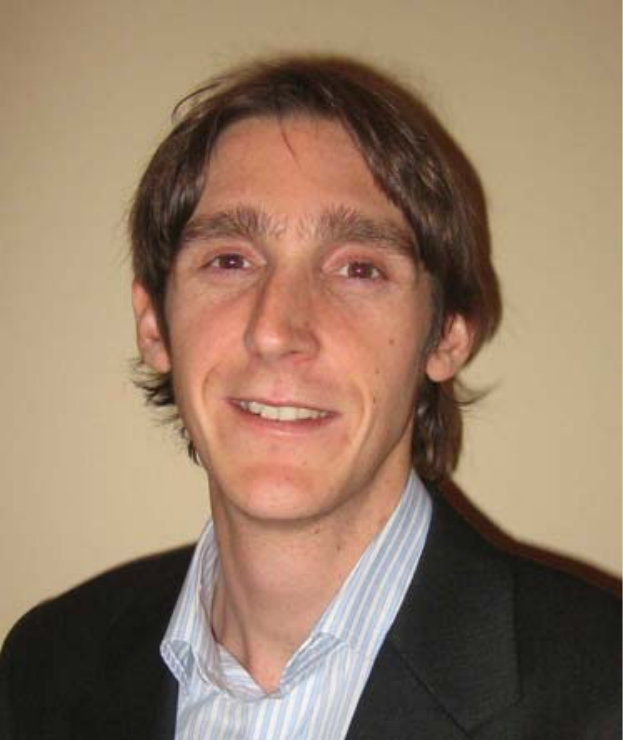}}]{Matthew R. McKay} (S'03-M'07) received the combined B.E. degree in Electrical Engineering and the B.IT. degree in Computer Science from the Queensland University of Technology, Australia, in 2002, and the Ph.D. degree in Electrical Engineering from the University of Sydney, Australia, in 2007. He then worked as a Research Scientist at the Commonwealth Science and Industrial Research Organization (CSIRO), Sydney, prior to joining the faculty at the Hong Kong University of Science and Technology (HKUST) in 2007, where he is currently the Hari Harilela Associate Professor of Electronic and Computer Engineering. He is also a member of the Center for Wireless Information Technology at HKUST, as well as an affiliated faculty member with the Division of Biomedical Engineering. His research interests include communications and signal processing; in particular the analysis and design of MIMO systems, random matrix theory, information theory, wireless ad-hoc and sensor networks, and physical-layer security.

Dr. McKay was awarded the University Medal upon graduating from the Queensland University of Technology. He and his coauthors have been awarded a Best Student Paper Award at IEEE ICASSP 2006, Best Student Paper Award at IEEE VTC 2006-Spring, Best Paper Award at ACM IWCMC 2010, Best Paper Award at IEEE Globecom 2010, Best Paper Award at IEEE ICC 2011, and was selected as a Finalist for the Best Student Paper Award at the Asilomar Conference on Signals, Systems, and Computers 2011. In addition, he received the 2010 Young Author Best Paper Award by the IEEE Signal Processing Society, the 2011 Stephen O. Rice Prize in the Field of Communication Theory by the IEEE Communication Society, and the 2011 Young Investigator Research Excellence Award by the School of Engineering at HKUST. Dr. McKay serves on the editorial boards of the IEEE Transactions on Wireless Communications and the mathematics journal, Random Matrices: Theory and Applications. In 2011, he served as the Chair of the Hong Kong Chapter of the IEEE Information Theory Society, whilst previously serving as the Vice-Chair and the Secretary. He has also served on the technical program committee for numerous international conferences, as well as the Publications Chair for IEEE SPAWC 2009, Publicity Chair for IEEE SPAWC 2012, and Poster Chair for IEEE CTW 2013.
\end{IEEEbiography}

\end{document}